\documentclass[twocolumn]{aastex63}

\newcommand{\kms}{{km s$^{-1}$}}

\newcommand{\magarcsec}{{mag arcsec$^{-2}$}}

\newcommand{\myemail}{jskang@astro.snu.ac.kr}
\newcommand{\profemail}{mglee@astro.snu.ac.kr}

\received{February 24, 2022}
\accepted{September 29, 2022}
\submitjournal{ApJ}

\shorttitle{Photometry of GCs in M104}
\shortauthors{Kang et al.}

\begin{document}

\title{ 
Tracing the Giant Outer Halo of 
the Mysterious 
Massive Disk Galaxy M104 \\
	I. Photometry of the Extended Globular Cluster Systems 
	\footnote{Based on observations obtained with MegaPrime/MegaCam, a joint project of CFHT and CEA/DAPNIA, at the Canada-France-Hawaii Telescope (CFHT) which is operated by the National Research Council (NRC) of Canada, the Institut National des Science de l'Univers of the Centre National de la Recherche Scientifique (CNRS) of France, and the University of Hawaii.}}

\correspondingauthor{Myung Gyoon Lee}
\email{\myemail,\profemail}

\author[0000-0003-3734-1995]{Jisu Kang}
\author[0000-0003-2713-6744]{Myung Gyoon Lee}
\affil{Astronomy Program, Department of Physics and Astronomy, SNUARC, Seoul National University, 1 Gwanak-ro, Gwanak-gu, Seoul 08826, Republic of Korea}

\author[0000-0002-2502-0070]{In Sung Jang}

\affiliation{Department of Astronomy \& Astrophysics, University of Chicago, 5640 South Ellis Avenue, Chicago, IL 60637, USA}

\author[0000-0001-6333-599X]{Youkyung Ko}
\affiliation{Korea Astronomy and Space Science Institute, 776 Daedeok-daero, Yuseong-gu, Daejeon 34055, Republic of Korea}

\author[0000-0002-9254-144X]{Jubee Sohn}
\affil{Astronomy Program, Department of Physics and Astronomy, SNUARC, Seoul National University, 1 Gwanak-ro, Gwanak-gu, Seoul 08826, Republic of Korea}

\author[0000-0002-2013-1273]{Narae Hwang}
\author{Byeong-Gon Park}
\affiliation{Korea Astronomy and Space Science Institute, 776 Daedeok-daero, Yuseong-gu, Daejeon 34055, Republic of Korea}

\begin{abstract}

M104 (NGC 4594, the Sombrero galaxy) is a mysterious  
massive early-type galaxy that shows a dominant bulge  and a prominent disk. 
However, the presence of a halo in M104 has been elusive, and it is not yet known how M104 has acquired such a peculiar structure. Using wide ($\sim2$ deg$^2$) and deep \textit{ugi} images of M104 obtained with the CFHT/MegaCam, we detect a large number of globular clusters (GCs) found 
out to 
$R\approx35'$ ($\sim100$ kpc). 
The color distribution of these GCs shows two subpopulations: 
a blue (metal-poor) system and a red (metal-rich) system. 
The total number of GCs is estimated to be $N_{GC}=1610\pm30$ and the specific frequency to be $S_{N}=1.8\pm0.1$.
The radial number density profile of the GCs is steep in the inner region at $R<20'$, and becomes shallow in the outer region at $20'<R<35'$.
The outer region is dominated by blue GCs and is extended out to $R\approx35'$. 
This shows clearly the existence of a giant metal-poor halo in M104. 
The inner region is composed of a bulge hosting a disk,
corresponding to a metal-rich halo as seen in early-type galaxies. 
At least two clumps of blue GCs are found in the outer region. 
One clump is overlapped 
with a faint stellar stream 
indicating that it may be a remnant of a disrupted dwarf galaxy. Our results imply that the metal-rich inner halo of M104 formed first via major mergers, and the  metal-poor outer halo grew via numerous minor mergers. 
\end{abstract}


\section{Introduction} \label{sec:intro}

\subsection{Mysterious Formation History of an Intriguing Disk Galaxy M104}

\begin{deluxetable*}{lccc}[htb!]
	\tabletypesize{\scriptsize}
	\tablecaption{Basic Parameters of M104 \label{table:info}}
	\tablehead{ 
		\colhead{Parameter} & 
		\colhead{Value} & 
		\colhead{Reference}}
	\startdata
	R.A.(J2000) 	& $12^h$ $39^m$ $59^s.4$ 	& NED	\\
	Decl.(J2000) 	& $-11\arcdeg$ $37\arcmin$ $23\arcsec$   & NED	\\
	Foreground extinction, $A_B$, $A_V$, $A_I$ & 0.185, 0.140, 0.077 & \citet{sch11} \\
	Foreground extinction, $A_u$, $A_g$, $A_i$ & 0.217, 0.169, 0.087 & \citet{sch11} \\
	Distance moduli, $(m-M)_0$ & $29.90\pm0.08$ & \citet{mcq16} \\
	Distance & $9.55\pm0.34$ Mpc & \citet{mcq16}	\\
	Image scale 
	 & 2.78 kpc arcmin$^{-1}$ & \citet{mcq16}	\\
	Total $B$-band magnitude, $B_T$	& $8.98\pm0.06$  & RC3 \\
	Extinction-corrected total $B$-band magnitude, $B_T^0$	& $8.39\pm0.06$  & RC3 \\
	Total $V$-band magnitude, $V_T$	& $8.00\pm0.06$  & RC3 \\
	Extinction-corrected total $V$-band magnitude, $V_T^0$	& $7.55\pm0.06$  & RC3 \\
	$B$-band absolute magnitude, $M_B$ & $-21.51\pm0.10$  & RC3, \citet{mcq16} \\
	$V$-band absolute magnitude, $M_V$ & $-22.35\pm0.10$  & RC3, \citet{mcq16} \\
	Position angle  & 90 deg ($B$), 88 deg ($K_s$) & RC3, 2MASS \\
	$D_{25}(B)$		& $522\farcs60\times214\farcs27$ ($8\farcm7\times3\farcm6$) & RC3 \\
	$D_{total}(K_s)$ & $594\farcs20 \times 320\farcs87$ ($9\farcm9\times5\farcm3$) & 2MASS \\
	Effective radius, $r_{\rm eff}$($V$) & $89''\pm2''$ ($4.1\pm0.1$ kpc)  & \citet{har14}, r$^{1/4}$ law \\
	 & $85''\pm5''$ ($3.9\pm0.2$ kpc)  & \citet{har14}, S\'ersic law \\
	Effective radius, $r_{\rm eff}$($I$) & $156\farcs2$ (7.2 kpc)  & \citet{jar11}, S\'ersic law \\
 	 & $117''\pm12''$ ($5.4\pm0.6$ kpc)  & \citet{jar11}, SB integration \\
	Heliocentric velocity, $v_{helio}$ & $1024\pm5$ \kms  & \citet{smi00} \\
	Central stellar velocity dispersion, $\sigma_0$ & $241.1\pm 4.4$ \kms & \citet{ho09} \\
	Maximum rotation velocity of stars, $v_{rot,s}$ & $229\pm 10$ \kms at $r\approx4'$ & \citet{ems96} \\
	Maximum rotation velocity of gas, $v_{rot,g}$ & $345$ \kms at $r=3\farcm2$  & \citet{sch78} \\
	Stellar mass, $M_*$ ($3.6 \mu m$) & $1.79\times10^{11} M_\odot$ & \citet{mun15} \\
	Stellar mass, $M_*$ ($K$) & $2.1\times10^{11} M_\odot$ & \citet{kar20} \\
	Dynamical mass, $M_{dyn}(R<41\rm kpc)$ & $1.3\times10^{12} M_\odot$ & \citet{dow14} \\
	Dynamical mass, $M_{dyn}(R<1 \rm Mpc)$ & $1.55\times10^{13} M_\odot$ & \citet{kar20} \\
	\hline
	Number of GCs, $N_{GC}$ & $1900\pm200$ & \citet{rho04} \\
	Specific frequency of GCs, $S_N$ & $2.1\pm0.3$ & \citet{rho04} \\
	Number of GCs, $N_{GC}$ & $1610\pm30$ & This study \\
    Specific frequency of GCs, $S_N$ & $1.8\pm0.1$ & This study \\
	\hline
	\enddata
\end{deluxetable*}

\begin{deluxetable}{lccc}[htb!]
	\tabletypesize{\scriptsize}
	\tablecaption{Morphological Type and Bulge-to-Total Ratio of M104 \label{table:type}}
	\tablehead{ 
		\colhead{Parameter} & 
		\colhead{Value} & 
		\colhead{Reference}}
	\startdata
	Type & SA(s)a	& RC3	\\
	 & Sa$^+$/Sb$^-$	& \citet{san94}	\\
	 & S0	& \citet{rho04}	\\
	 & E	& \citet{gad12}	\\
	 & (R)SA(l)0$^+$sp  & \citet{but15}	\\
	 & E(d)1-2	& \citet{but15}	\\
	 & E/S0	& Adopted in this study	\\
	 \hline
	B/T & 0.86 & \citet{ken88} \\
	 & 0.73 & \citet{jar11} \\
	 & 0.93 & \citet{kor11} \\
	 & 0.77 (BD\tablenotemark{a}) & \citet{gad12} \\
	 & 0.13 (BDH\tablenotemark{a}) & \citet{gad12} \\
	\hline
	\enddata
	\tablenotetext{a}{BD: Bulge+Disk model, BDH: Bulge+Disk+Halo model}
\end{deluxetable}

\begin{figure*}[htb!]
    \centering
    \includegraphics[scale=0.9]{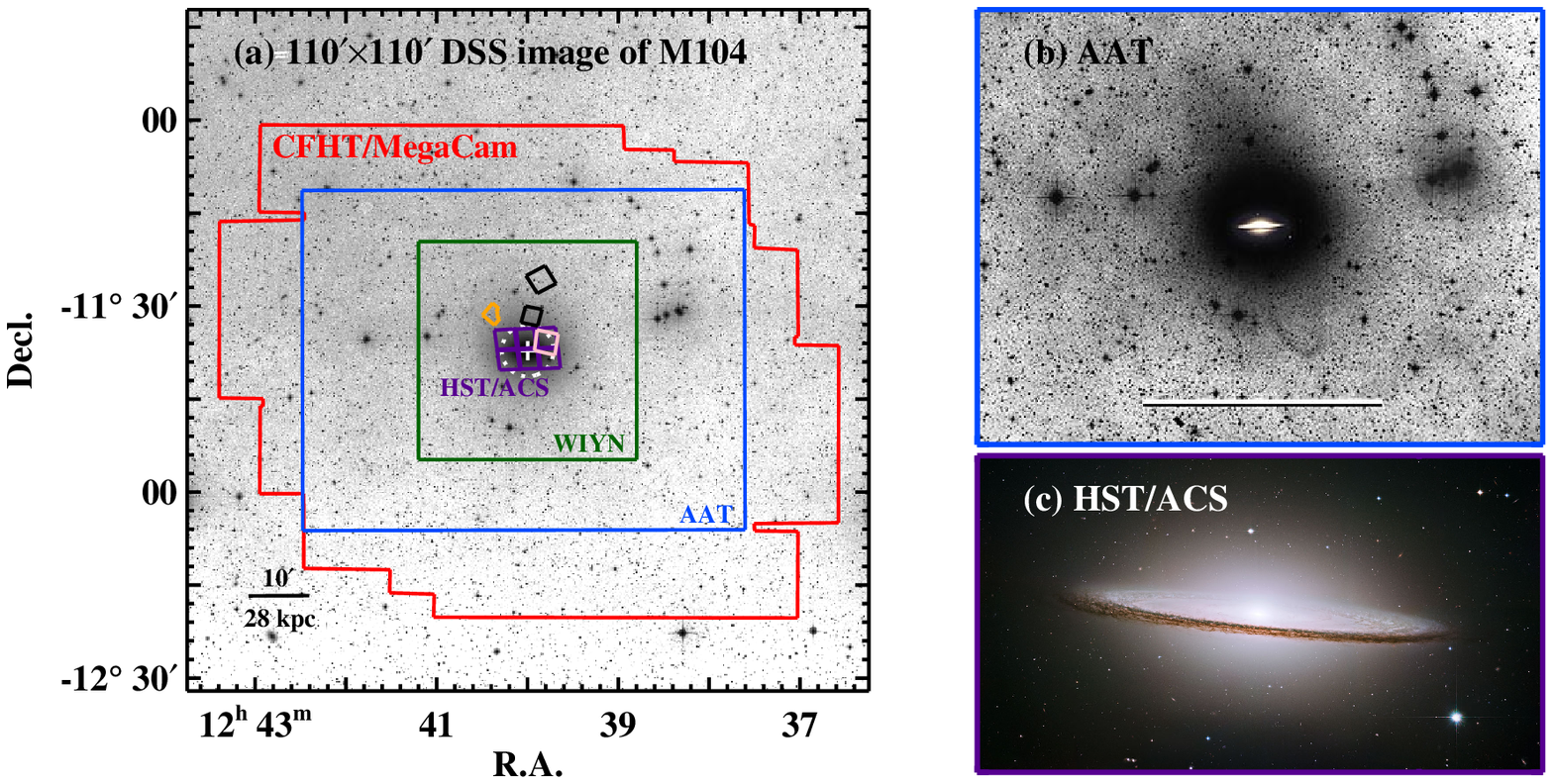}
	\caption{
		(a) $110'\times110'$ Digitized Sky Survey image centered on M104. 
		Red lines mark the CFHT/MegaCam coverage used in this study.
        Blue lines mark the location of David Malin's Anglo-Australian Telescope (AAT) deep image shown in (b).
        Green lines mark the WIYN field used in \citet{rho04} and \citet{har14}. 
		Purple lines mark the HST/ACS field (PI: Noll, PID: 9714) used in \citet{spi06} and \citet{har10}.
		The ACS color image is shown in (c).
		The WFPC2 field (PI: Griffiths, PID: 5369) used in \citet{lar01b} and \citet{mou10}   
		is marked in orange. 
		The ACS field (PI: McQuinn, PID: 13804) used in \citet{mcq16} is marked in pink. 
		The ACS and UVIS fields (PI: Goudfrooij, PID: 14175) used in \citet{coh20} are marked in black.
		White cross marks the center of M104, and white dotted circle indicates a circle with a diameter of $D_{25}$.
		North is up, and east is to the right. 
		The scale bar is marked on the bottom left side.
	}
	\label{fig:finding}
\end{figure*}

M104 (NGC 4594, the Sombrero galaxy) 
is one of the most massive nearby disk galaxies, 
and it is located 
at a distance of 9.55 Mpc 
\citep{mcq16}.
It is known as the brightest member of a fossil group where the $K$-band magnitude gap between the second and the first brightest galaxies is larger than two 
\citep[$\Delta K_{12} = 2.98$ mag,][]{mak11}.
According to the study of \citet{kar20}, M104 hosts 27 group members based on distances or radial velocities. 
Most of the group members are dwarf galaxies \citep{hau09,jav16,car20}. Fifteen of the group members have measured radial velocities, and the farthest of them is located at a projected distance of 
970 kpc (5.85\arcdeg). 
With these measurements, 
\citet{kar20} estimated the total group mass 
to be $(1.55\pm0.49)\times10^{13} M_\odot$. They also derive a high value for the virial mass to stellar mass ratio, $M_T/M_*=65\pm20$, which 
means 
that M104 is dominated by dark matter.  
M104 is located almost in an isolated environment in the sky, 
as the nearest large member galaxies are located farther away at a projected separation of $r_p>2.2\arcdeg$. 

Table \ref{table:info} 
lists some basic parameters of M104.

M104 is well-known 
mostly 
for its unique morphology. 
It has a dark dust lane and a dominant spheroidal component as shown in Figure \ref{fig:finding}(c)\footnote{\url{https://esahubble.org/images/opo0328a/}}. 
Due to 
the unusual morphology, 
M104's 
morphological classification 
has been 
complicated and controversial. 
Table \ref{table:type} summarizes the morphological type of M104 
given in previous studies. 
Conventionally, M104 is classified as 
Sa to Sb because of its prominent disk and dust lane \citep{dev91,san94}. 

However, more and more evidence suggest that M104 is actually an elliptical or a lenticular galaxy rather than a spiral galaxy \citep{rho04,gad12,but15}. First, M104 is massive \citep[stellar nass $M_*=1.79\times10^{11} M_\odot$,][]{mun15}
with a large rotation velocity of the gas in the central region at $R\sim3\farcm2$ \citep[$v_{rot,g}\approx350$ \kms,][]{sch78}, 
and is as bright as M86, a giant elliptical galaxy in the Virgo cluster ($M_{V}\approx-22.4$ mag).
Second, its bulge fraction is extremely 
large compared with other spiral galaxies, 
showing a bulge-to-total ratio of B/T$=0.73$ to 0.93 
\citep[][see Table \ref{table:type}, 
although \citet{gad12} derived a much lower value, B/T=0.13, for the model considering three components: a bulge, a disk, and a halo]{ken88,jar11,kor11,gad12}. 
Third, M104 hosts a much larger number of globular clusters (GCs) compared to other spiral galaxies 
\citep[$N_{GC}=1900\pm200$, specific frequency $S_{N}=2.1\pm0.3$,][]{rho04},
which 
 led \citet{rho04} to 
classify M104 as an S0 galaxy. 
Fourth, the metallicity distribution of the resolved stars in the outer region (at $R=33$ kpc) of M104 
shows a dominant population of metal-rich stars ($[Z/H]>-0.3$ dex) and a negligible fraction of metal-poor stars, which is similar to that of the massive elliptical galaxy NGC 5128 
\citep{mou10,coh20}.
Using the Spitzer 3.6$\mu$m images of M104, \citet{but15} 
found a well-defined outer ring and a smooth bright inner disk, 
but they found no clear spiral structure and classified M104 as (R)SA(l)0$^+$sp or E(d)1-2.

For these reasons, M104 is considered 
a mysterious early-type galaxy (ETG) with a 
prominent disk. 
Therefore, it is worth investigating if M104 had experienced a morphological transition 
and how it had acquired such a massive disk in a low-density environment.

\subsection{Expectations on the Giant Halo around M104}

Since M104 is a massive ETG, it is expected that there is a giant stellar halo around M104. 
This, combined with it's distinctive feature of being a massive edge-on galaxy located in an isolated environment, makes M104 an ideal target to study a galaxy's outer halo. 
According to the current paradigm of $\Lambda$CDM cosmology, massive ETGs form and grow 
via hierarchical merging of less massive galaxies and accretion of dwarf satellite galaxies. 

The existence of a stellar halo around M104 has been expected in previous studies with deep surface photometry. 
First, \citet{bec84} found a faint halo at the galactocentric distance of $R\approx9'$ of M104 
from deep IIIa--F plate imaging
(similar to the R-band imaging).  
The surface brightness of the halo of M104 reaches $\mu_{R}\approx$ 25 \magarcsec, 
and the color is redder than the color of other disk galaxies 
such as NGC 253 ($(B-V)_T\approx1.0$).
\citet{bur86} also found an ellipsoidal component of M104 with similar size and color to that of \citet{bec84}, and concluded that it is a halo. 
Later, \citet{mal97a} and \citet{mal97b} also found a more extended faint halo from much deeper and blue-sensitive  IIIa--J plate imaging
(similar to $B$-band imaging), 
reaching $\mu_{B}\approx$ 28 \magarcsec~at $R\approx20'$. 
They found, for the first time, 
faint substructures 
such as 
 the south-west `loop' and 
 the north-east `fan' as shown in Figure \ref{fig:finding}(b)\footnote{\url{http://www.messier.seds.org/more/m104_deep.html}}.
These substructures were re-examined 
recently by \citet{mar21}. 
Lastly, based on 
detailed  structural analysis of the \textit{Spitzer} IRAC 3.6$\mu m$ image of M104,
\citet{gad12} 
found that the surface brightness profile of M104 fits better 
with the model of a pseudo-bulge 
with a disk embedded in an exponential halo ($R_e\approx2'-3'$), 
than with the model of a classical bulge 
with a disk component. 
They also suggested a scenario where M104 is an elliptical galaxy which acquired a disk later.


However, 
these studies could not trace the giant metal-poor halo of M104. 
The halo expected from the previous studies is rather small and too red (metal-rich), 
indicating that 
they 
are still missing the giant metal-poor halo.
For this reason, we 
focus mainly on the outskirts of M104 in order to trace the outer metal-poor halo of M104 and study its assembly history. 

\subsection{Globular Clusters as Useful Tracers of the Giant Halo around M104}

In this study, we utilize GCs to trace the giant outer halo around M104. 
In general, extragalactic GCs are useful to study the assembly history of their host galaxies for several reasons \citep{ash98,bro06}.
First, they have survived for a Hubble time and are age-datable, 
making them excellent tracers of formation and evolution history of their host galaxies.
Next, GCs are bright and compact, 
so they are found in every region of galaxies 
and are easily detected even in faint stellar halos.
Third, GCs are important dynamical tracers for ETGs like M104 
which often contain 
a large number of GCs. 
One of the most important  properties of GCs is that they show a bimodality in 
their color distributions. 
Blue and red colors of 
GCs 
correspond to metal-poor and metal-rich subpopulations, 
assuming that the ages of the GCs are as old as the age of Milky Way GCs.
These subpopulations are distinguishable in various properties,
and not just in color distributions. 
Therefore, we can trace the different components, 
such as the outer halo, of host galaxies by utilizing GCs.


There are several studies dealing with the photometry of the GCs in the central region of M104 \citep{lar01b,rho04,spi06,har10,har14}. 
However, 
 these previous studies of M104 GCs covered only the inner region 
 at $R<19'$ ($\approx50$ kpc), as shown in Figure \ref{fig:finding}(a). 
Consequently, spectroscopic studies were also limited to $R<19'$ \citep{dow14}. 
This coverage is not wide enough to study the outer halo of M104 and is much smaller than 
the stellar halo of Centaurus A \citep{crn16} or Andromeda galaxy \citep{mcc09} which extends out to around 150 kpc. 
Therefore, additional observations covering the outer field of M104 is needed 
to trace the giant outer halo.
In this study, we present a wide-field photometric survey of the GCs in M104 
covering about 2 deg$^2$ ($100'\times75'$ $\approx$ 280 kpc $\times$ 200 kpc) field of view using the MegaCam mounted on the 3.6 m Canada-French-Hawaii telescope (CFHT).

\begin{deluxetable*}{cccccccc}[htb!]
	\tabletypesize{\scriptsize}
	\tablecaption{A Summary of CFHT/MegaCam Observations	\label{table:obs}}
	\tablehead{ 
		\colhead{Object} &
		\colhead{Proposal ID} & 
		\colhead{Date UT} & 
		\colhead{Band} & 
		\colhead{Exposure Time} &
		\colhead{Seeing} &
		\colhead{Airmass} &
		\colhead{Background} }
	\startdata
	&        & 2015 May 17 & $u$
	& 
	285s$\times7=1995$s & 0\farcs8 & 1.21 & \\
	M104 & 15AK06 & 2015 May 24 & $g$
	& 
	598s$\times8=4784$s & 0\farcs8 & 1.19 & Dark \\
	&        & 2015 May 25 & $i$
	& 
	800s$\times7=5600$s & 0\farcs5 & 1.18 & \\
	\hline
	\enddata
\end{deluxetable*}

This paper is organized as follows. 
We describe how we obtained and reduced our data in Section \ref{sec:data}.
We identify the GC candidates and investigate their photometric properties in Section \ref{sec:results}.
We compare our results with previous studies 
and discuss them in regards with the formation and evolution of M104 in Section \ref{sec:discuss}.
Finally, we summarize our results in Section \ref{sec:summary}. 
We adopt the distance of 9.55 Mpc ($(m-M)_{0}=29.90\pm0.08$) 
to M104 based on the tip of the red giant branch (TRGB) method given by \citet{mcq16}.
At this distance, 
one arcminute corresponds to 2.78 kpc. 

\section{Observations and Data Reduction} \label{sec:data}

\subsection{CFHT/MegaCam Observations}

We obtained wide and deep \textit{ugi} images of M104 
with CFHT/MegaCam as 
part of the K-GMT Science Program (PI: Myung Gyoon Lee, PID: 15AK06). 
Table \ref{table:obs} shows a summary of observations. 
The CFHT is a 3.6m optical/infrared telescope located at Mauna Kea, 
and the MegaCam is a wide-field imaging camera covering a $1\arcdeg \times 1\arcdeg$ field
with a pixel scale of 0\farcs185. 
The camera consists of 40 mosaic CCD chips, 
so we chose a large dithering pattern to fill the largest gaps in the mosaic 
and to avoid artifacts such as internal reflection in the camera optics. 
We obtained 7, 8, and 7 shots, 
respectively, for 
the \textit{u, g, i}-bands
so that the total exposure times are 
about 2000s, 4800s, and 5600s. 
The MegaCam coverage marked in Figure \ref{fig:finding} is the result of seven different pointings with a large dithering pattern. 
These three filters were chosen to select GC candidates effectively \citep[e.g.,][]{lim17,ko19}. 
The data was obtained under good conditions with a seeing of $0\farcs5$ in \textit{i}-band, an airmass of 1.2, and a dark sky background.

\subsection{Image Combining}
In order to obtain wide and deep images in each filter, 
multiple shots taken with the same filters are combined with the following steps. 
First, we create a weight image for each chip to mask bad or saturated pixels. 
Second, we create source catalogs using \texttt{Source Extractor} \citep{ber96} with a detection threshold of 3$\sigma$ for \textit{u}-band and 5$\sigma$ for \textit{g} and \textit{i}-bands. The weight images are used when running \texttt{Source Extractor}.
Third, astrometric solutions are found using \texttt{SCAMP} \citep{ber06}. 
We select 
the Pan-STARRS catalog for our reference. 
The FWHM range is set from 2 to 10 pixels for \textit{u} and \textit{g}-bands and 2 to 6 pixels for \textit{i}-band. 
In the case of \textit{u} and \textit{g}-bands, using the solutions found in \textit{i}-band as \texttt{*.ahead} input file led to better results.
CCD chips that cannot find proper astrometric solutions are removed after visual inspection. 
Finally, background-subtracted deep and wide images are obtained by median-combining all the chips using \texttt{SWarp} \citep{ber02}. 
Weight images are also used when running \texttt{SWarp}. 

In Figure \ref{fig:finding}(a) we mark the coverage of our CFHT/MegaCam images after combining all the shots. 
Note that the coverage of this study, about 2 deg$^2$ ($100'\times75'$ $\approx$ 280 kpc $\times$ 200 kpc), is 
more than 5 times wider than the KPNO/WIYN data used in \citet{rho04} and \citet{har14}. 
The WIYN data is the widest GC survey 
of M104 in the previous studies. 
The maximum galactocentric radius of our data is $R=54$ arcmin ($\approx$ 150 kpc).

\subsection{Photometry and Calibration} \label{sec:cal}

We perform source detection and aperture photometry for the combined images 
using \texttt{Source Extractor} \citep{ber96}. 
The \textit{i}-band image is used as a reference image to detect sources in each image with a detection threshold of 3$\sigma$. 
The FWHM value is 4.4 pixels 
(0\farcs8) 
for 
the \textit{u} and \textit{g}-band images and 2.4 pixels 
(0\farcs5) 
for the \textit{i}-band image. 
We obtain the aperture magnitudes with various aperture sizes to derive magnitude concentration parameters and to calibrate the photometry as described in the next paragraph. 

Photometric calibration is conducted using bright SDSS stars \citep{ahu20}. 
The SDSS catalog provides various kinds of magnitudes, and we use `psfMag' (PSF magnitude) because 
it has the smallest calibration error compared to using other types of magnitudes such as `fiberMag'. 
From the SDSS DR16 catalog we get the coordinates and \textit{ugri} PSF magnitudes of the bright SDSS stars in the M104 field. 
The magnitudes are then  transformed to the MegaCam system according to the following relation\footnote{\url{http://www.cadc-ccda.hia-iha.nrc-cnrc.gc.ca/en/megapipe/docs/filtold.html}}:

$u_{Mega} = u_{SDSS} - 0.241 (u_{SDSS} - g_{SDSS})$,

$g_{Mega} = g_{SDSS} - 0.153 (g_{SDSS} - r_{SDSS})$, and

$i_{Mega} = i_{SDSS} - 0.003 (r_{SDSS} - i_{SDSS})$.

Magnitude differences are calculated 
between the MegaCam magnitudes that were transformed from the SDSS PSF magnitudes and the aperture magnitudes obtained in this study: 

$u (r=9{\rm pix})-u_{Mega}=0.04 \pm 0.13$, 

$g (r=9{\rm pix})-g_{Mega}=0.05 \pm 0.05$,  and

$i (r=5{\rm pix})-i_{Mega}=-0.04 \pm 0.06$.

When calculating the magnitude difference, we choose 
a 9 pixel radius aperture in 
the \textit{u} and \textit{g}-band 
images and 
a 5 pixel radius aperture in 
the \textit{i}-band image 
because these led to the smallest error. 
We use CFHT AB magnitude system in this study. 

Due to the wide coverage of the MegaCam, 
the PSF 
can vary according to the CCD positions within a single shot. 
Moreover, the images of the point sources can be broadened 
after image combining with imperfect astrometric solutions. 
For these reasons, we tried to perform PSF photometry for the individual CCD chips 
to resolve GCs more carefully and clearly,
 but decided to use a different method for two reasons.
First, there were not enough PSF stars within the field of view of each chip so PSF modeling was difficult.
Second, photometric calibration for each chip was impossible 
because 
the SDSS covers only a part of the M104 field. 
Therefore, we decide to obtain photometric results from aperture photometry of the combined images and calibrate them with the PSF magnitudes of the SDSS stars.


\subsection{Completeness Test}

\begin{figure}[b!]
    \centering
	\includegraphics[scale=0.9]{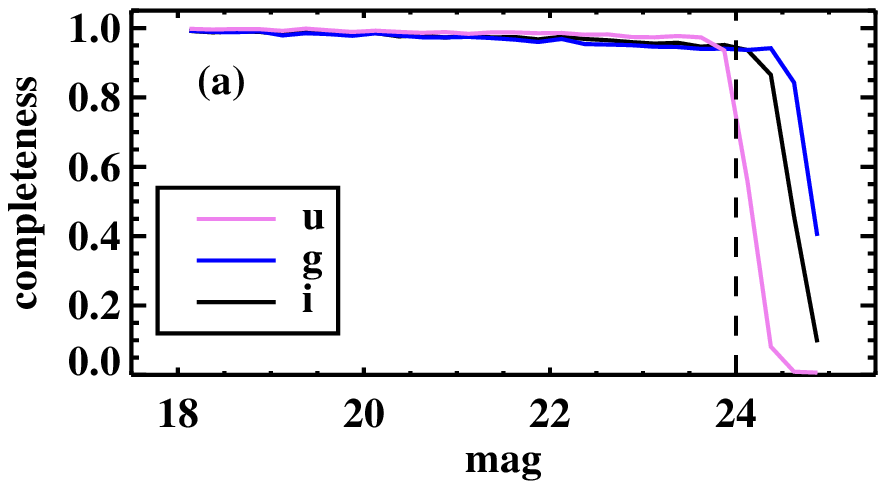}
	\includegraphics[scale=0.9]{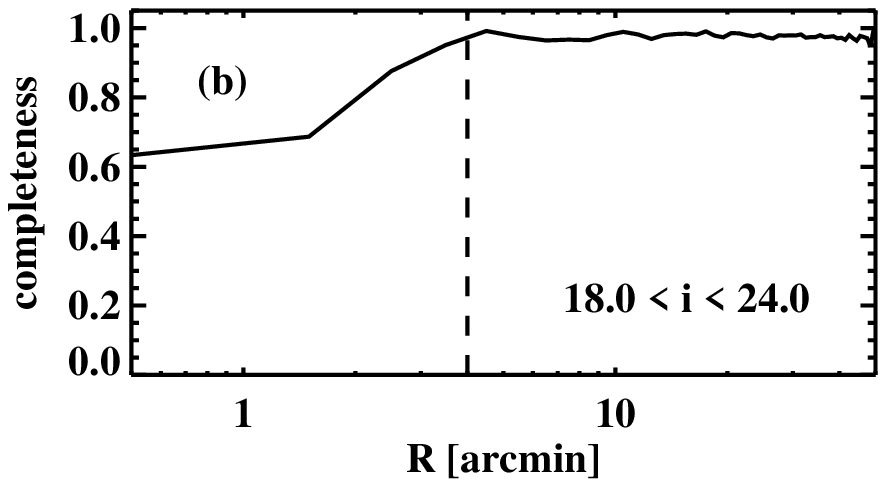}
	\caption{(a) Completeness as a function of 
 magnitude. 
 Pink, blue, and black lines mark $u$, $g$, and $i$-bands, respectively. The completeness is over 95\% for bright sources with $i<24$ mag.  
	(b) Completeness for bright sources with $18<i<24$ mag as a function of galactocentric distance. The completeness is over 95\% at $R>4\arcmin$.
	}
	\label{fig:comp}
\end{figure}

To check the magnitude limit of the image and to check the incompleteness in the central region, we perform an artificial star test. 
We inject 5,000 artificial stars with $18<i<25$ mag and a FWHM value of 2.4 pixels in the \textit{i}-band combined image. 
We repeat this for 10 images and calculate the recovery rate for the 50,000 stars in total. 
Figure \ref{fig:comp}(a) shows the completeness along the \textit{i}-band magnitude, showing that the completeness is over 95\% for the sources with $i<24$ mag and over 50\% for the sources with $i<24.5$ mag. 
Figure \ref{fig:comp}(b) shows the completeness for the bright sources with $18<i<24$ mag along the galactocentric distance. We can see that the completeness is over 95\% at $R>4\arcmin$. 
We repeat the test for the $g$ and $u$-band images and present the result in Figure \ref{fig:comp}(a). 
For the $g$-band, the completeness is 90\% at 24.5 mag and 50\% at 24.8 mag. For the $u$-band, the completeness is 90\% at 23.9 mag and 50\% at 24.2 mag. 

\section{Results} \label{sec:results}

\subsection{Selection of GC Candidates}

We select GC candidates using the size (Figure \ref{fig:cindex}), color (Figure \ref{fig:ccd}), and magnitude (Figure \ref{fig:cindex2}) of the detected sources, 
and then apply the galactocentric distance limit (Figure \ref{fig:rdp0})
to this sample for further analysis. 
As a result, we find 2936 GC candidates in total.
Table \ref{table:cat} lists a small portion of the entire catalog of GC candidates.

\begin{deluxetable*}{ccccccc}[htb!]
	\tabletypesize{\scriptsize}
	\tablecaption{A $ugi$ Photometric Catalog of the GCs in M104	\label{table:cat}}
	\tablewidth{0pt}
	\tablehead{
		\colhead{ID} & 
		\colhead{R.A. (J2000)} & 
		\colhead{Decl. (J2000)} & 
		\colhead{$i$} & 
		\colhead{$g-i$} & 
		\colhead{$u-g$} &
		\colhead{$C_i$} \\
		\colhead{} &
		\colhead{(deg)} &
		\colhead{(deg)} &
		\colhead{(mag)} &
		\colhead{(mag)} &
		\colhead{(mag)}
	}
	\startdata
0001 & 189.40932 & -11.58356 & $21.76 \pm 0.06$ & $0.95 \pm 0.08$ & $1.74 \pm 
0.22$ & $0.39$ \\
0002 & 189.40988 & -11.57188 & $18.88 \pm 0.06$ & $0.67 \pm 0.08$ & $0.92 \pm 
0.14$ & $0.40$ \\
0003 & 189.41223 & -11.67494 & $22.36 \pm 0.06$ & $1.03 \pm 0.09$ & $1.69 \pm 
0.35$ & $0.38$ \\
0004 & 189.41660 & -11.71237 & $19.66 \pm 0.06$ & $0.97 \pm 0.08$ & $1.68 \pm 
0.14$ & $0.36$ \\
0005 & 189.41745 & -11.69852 & $22.26 \pm 0.06$ & $1.05 \pm 0.09$ & $1.63 \pm 
0.31$ & $0.39$ \\
	\hline
	\enddata
	\tablecomments{This table is published in its entirety in the electronic edition. The five samples are shown here as a guide for the table content.
    }
\end{deluxetable*}

\subsubsection{Size Distribution of Detected Sources}

\begin{figure}[tb!]
    \centering
	\includegraphics[scale=0.9]{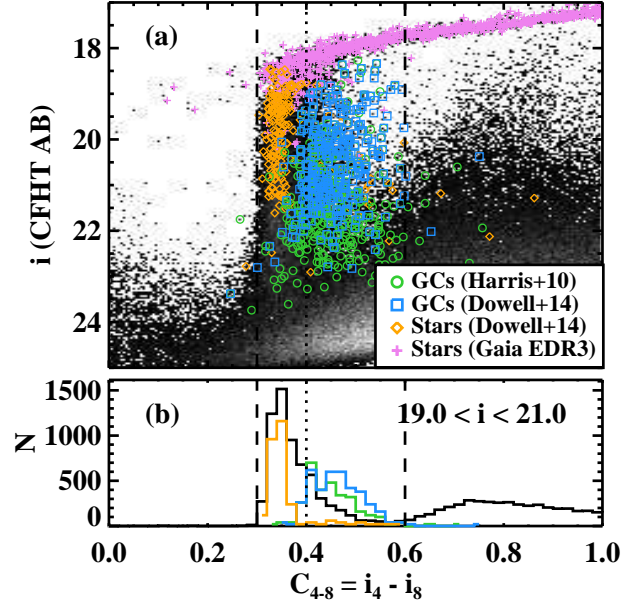}
	\caption{(a) $i$-band magnitude vs. C-index diagram of the sources 
		with $i < 25$ mag. 
        We overplot the GCs 
        \citep{har10} in green circles, GCs and foreground stars 
        \citep{dow14} in blue squares and yellow diamonds, and foreground stars 
        \citep{gai21} in pink plus symbols.
		(b) The distribution of the C-index of bright sources with $19 < i < 21$ mag (black lines). 
		The histograms of confirmed GCs (cyan and green lines) and foreground stars (yellow lines) are exaggerated 20 times.
		We select GC candidates with $0.3 < C < 0.6$. 
	}
	\label{fig:cindex}	
\end{figure}

First, we use the magnitude concentration parameter (C-index) as a proxy for size to select GC candidates. 
The C-index is defined by the difference between the magnitude with small aperture size and the magnitude with large aperture size. 
This method quantifies how the brightness profile is centrally concentrated.
Small C-index values correspond to narrow point-like profiles 
and large C-index values correspond to broad extended profiles. 
The seeing condition (FWHM=2.4 pixels
=0\farcs5
) is the best for the \textit{i}-band image so we use the \textit{i}-band image for deriving C-index values of the detected sources.
At the distance to M104, 1 arcsecond corresponds to 46.3 pc, so the best seeing condition of 0.5 arcsecond corresponds to $\sim23$ pc. Effective radii (diameters) of typical GCs are about 3 pc (6 pc), which corresponds to 0\farcs06 (0\farcs12) at the distance of M104. So the images of the GCs will appear as slightly extended sources and point sources in the CFHT images, which can be distinguished by the C-index.
In this study, we define $C=C_{4-8} = i_4 - i_8$ 
where the small aperture diameter corresponds to 4 pixels (0\farcs74) and the large aperture diameter corresponds to 8 pixels (1\farcs48). 

Figure \ref{fig:cindex}(a) is the density plot of \textit{i}-band magnitude 
versus C-index of the sources with $i < 25$ mag. 
As a reference for setting the selection criteria of GC candidates, we use the lists of GCs and foreground stars found in previous studies. 
In this figure, 
green symbols mark the GCs that were resolved in the HST/ACS images \citep{har10},
and blue and yellow symbols mark the GCs and the foreground stars that were spectroscopically confirmed \citep{dow14}. 
We also mark the foreground stars selected from the Gaia parallax with pink symbols \citep{gai21}.

Several features are noticeable in Figure \ref{fig:cindex}(a).
First, the foreground stars from Gaia are mostly brighter than $i=19$ mag and are distributed in a wide range of C-index. This is because these sources are saturated 
in the CFHT images used in this study.
Recent studies including \citet{vog20} and \citet{hug21} utilized Gaia data to select GCs around Cen A. In this study, however, we did not use Gaia data to select GCs because the magnitude range of Gaia data only covers the bright end of M104 GCs and they are mostly saturated. 
Second, the foreground stars from \citet{dow14} are mostly brighter than $i=21$ mag and are located in a narrow range of C-index, $0.3<C<0.4$. 
From this, we can expect 
foreground stars (point sources) 
to be located in this range of C-index. 
Third, the GCs from \citet{har10} and \citet{dow14} are mostly brighter than $i=23$ mag and are located in the range of $0.4<C<0.6$. 
This means that the GCs are more extended than the point sources in the image so they are resolved as compact sources.
It is possible to distinguish GCs from the foreground stars with C-index 
because of the proximity of M104 and the high resolution of our data. 
However, as 
magnitudes become fainter, some GCs are located in a C-index range where point sources are expected to be located. 
Therefore, we select the sources with $0.3<C<0.6$ as GC candidates, including both point sources ($0.3<C<0.4$) and compact sources ($0.4<C<0.6$). 
The sources with $C<0.3$ would be artifacts 
and the sources with $C>0.6$ would be background galaxies. 

Figure \ref{fig:cindex}(b) is a histogram of the C-index distributions of bright (but not saturated) sources with $19<i<21$ mag (black lines).
The histograms of the GCs and foreground stars from other studies are marked with the same color as above after 20 times exaggeration. 
It is clear that GCs and foreground stars are separated very well in this bright magnitude range. 

\subsubsection{Color-color Diagram of Point/Compact Sources}

\begin{figure}[b!]
    \centering
	\includegraphics[scale=0.9]{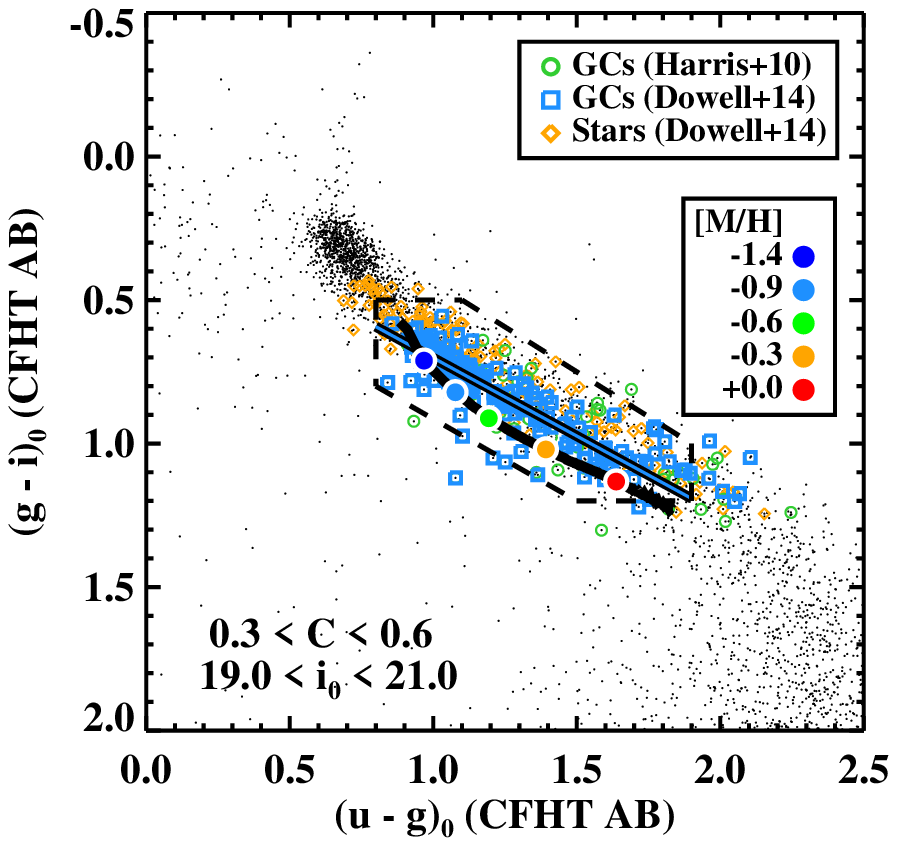}
	\caption{ 
        $(g-i)_0-(u-g)_0$ color-color diagram of bright point/compact sources with $0.3<C<0.6$ and $19< i_0 < 21$ mag.
        We also overplot the GCs \citep{har10,dow14} and foreground stars \citep{dow14}.
        Black dashed hexagon marks the GC selection criteria. 
        Blue solid line marks the color-color relation derived from linear fitting for the GCs. 
        Black solid line marks the relation derived from the 12 Gyr PARSEC stellar models \citep{bre12}. 
        Filled circles mark the [M/H] values. 
	}
	\label{fig:ccd}
\end{figure}

Second, we use colors to select GC candidates. 
Figure \ref{fig:ccd} is the $(g-i)_0-(u-g)_0$ color-color diagram of the bright point/compact sources with $19<i_0<21$ mag and $0.3<C<0.6$.
The color and magnitude with subscript 0 represent foreground extinction-corrected values.
Foreground extinction was corrected with the value of $A_u$, $A_g$, $A_i=$ 0.217, 0.169, 0.087, respectively \citep{sch11}. 
As in Figure \ref{fig:cindex}, we also mark the GCs and foreground stars from \citet{har10} and \citet{dow14}, and use them to set the criteria for the selection of the GC candidates.
This color-color combination is well-known to effectively distinguish GCs from other objects.
For example, \citet{lim17} used $(g-i)_0-(u-g)_0$ color-color diagram to select GC candidates in NGC 474 in CFHT/MegaCam images. 
\citet{ko19} also used the same criteria to select GC candidates in M85. 
With the same selection criteria as in \citet{lim17} (see their Figure 2) and \citet{ko19} (see their Figure 5), most of the previously known GCs in M104 are also included. 
Therefore, we adopt the same selection criteria for GC candidates following the previous studies.


\begin{figure}[b!]
    \centering
	\includegraphics[scale=0.9]{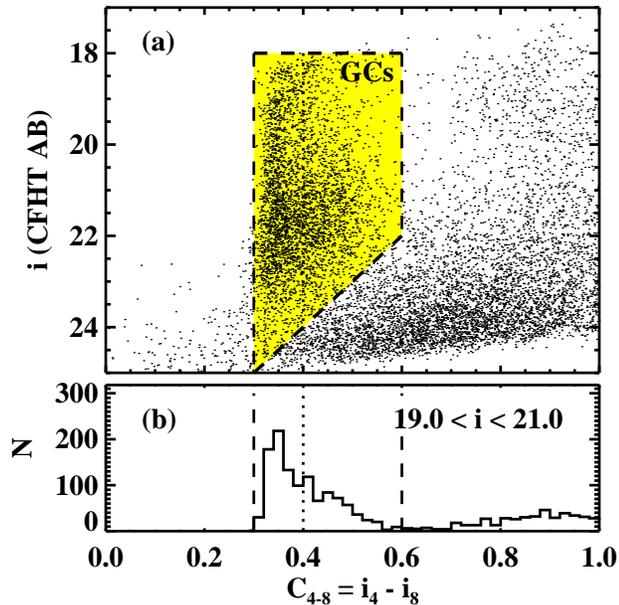}
	\caption{
	    Same as Figure \ref{fig:cindex} but for sources with $i < 25$ mag and GC-like colors (within the dashed hexagon marked in Figure \ref{fig:ccd}).
		To remove the background galaxy contamination, we set the faint magnitude range as marked by the black dashed diagonal line.
	    }
	\label{fig:cindex2}
\end{figure}

We fit the color-color relation linearly for the previously known GCs, and obtain the relation as follows (blue line):
$(g-i)_0 = (0.54\pm0.02)(u-g)_0 + (0.16\pm0.02)$ (RMS=0.09). 
We also overlay 
the 12 Gyr PARSEC isochrones for [M/H] = $-2.2$ to $+0.3$ \citep[black line,][]{bre12}. 
The color-color relation and the color ranges of the isochrones are very similar to the empirical one.

\subsubsection{Magnitude Range of GC Candidates}

Third, we set the magnitude range to select GC candidates. 
Figure \ref{fig:cindex2} again shows \textit{i}-band magnitude versus C-index of the sources with $i < 25$ mag as shown in Figure \ref{fig:cindex}. 
This time we only plot the sources satisfying the color criteria marked in Figure \ref{fig:ccd}. 
As the magnitude becomes fainter, the number of sources increases and becomes maximum at $i\approx22$ mag and decreases to $i\approx24$ mag and then increases again. 
The magnitude where the number of GCs becomes maximum is called the GC luminosity function (GCLF) turnover magnitude (TOM), and we will discuss it in Section \ref{sec:gclf}. 
The increasing number of faint sources with $i\gtrsim24$ mag are mostly background sources.
Therefore, to exclude background contamination, we select only the sources brighter than $i=(28-10C_{4-8})$ mag as our final GC candidates.
We also select sources fainter than $i=18$ mag to avoid saturated sources. 


\subsubsection{Radial Distribution of GC Candidates}

\begin{figure}[tb!]
    \centering
	\includegraphics[scale=0.9]{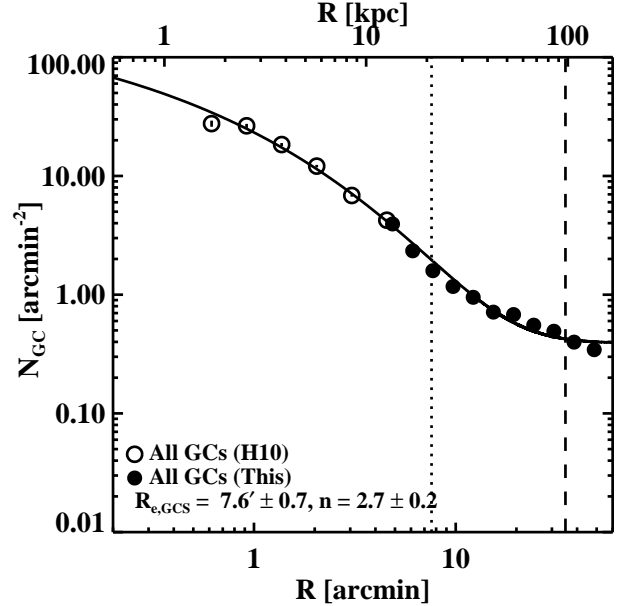}
	\caption{
	    Radial number density profile of the GC candidates according to galactocentric distance. 
        We use the GCs found in the HST/ACS images by \citet{har10} (open circles) and the GC candidates from this study (filled circles) together to construct a complete profile from the center to the outer region. 
        The solid line marks the fitted S\'ersic profile,  
	    the dotted line marks the effective radii of the GC systems, and the dashed line marks the radial coverage limit of the GCs.
    }
	\label{fig:rdp0}
\end{figure}

\begin{figure*}[htb!]
    \centering
	\includegraphics[scale=0.65]{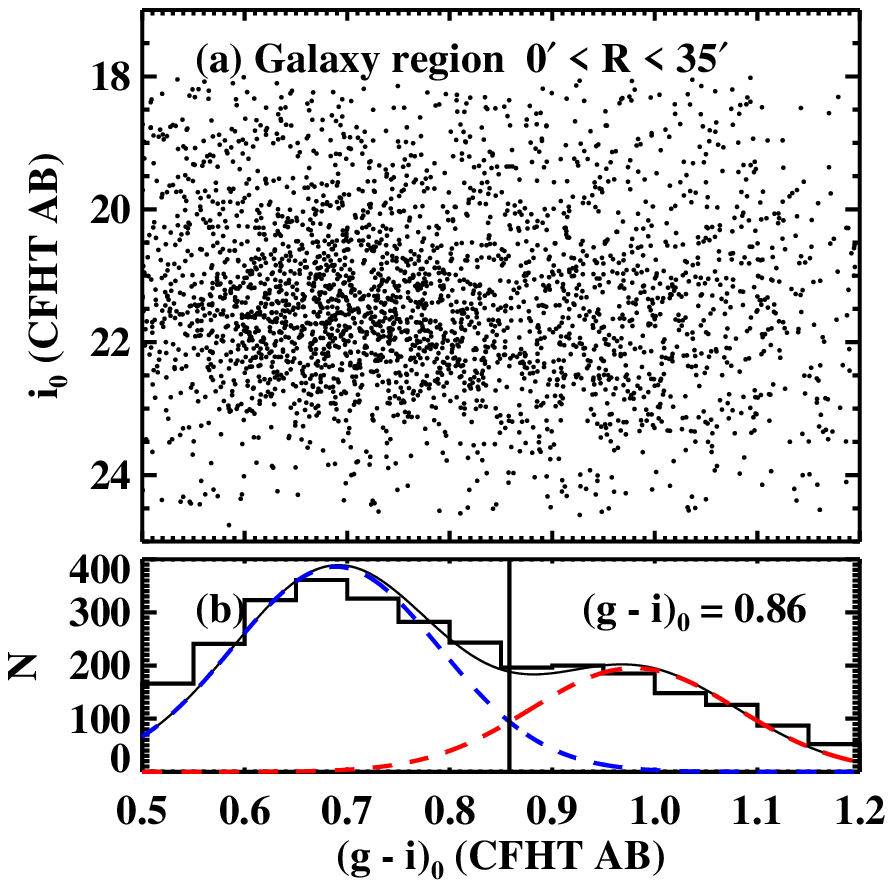}
	\includegraphics[scale=0.65]{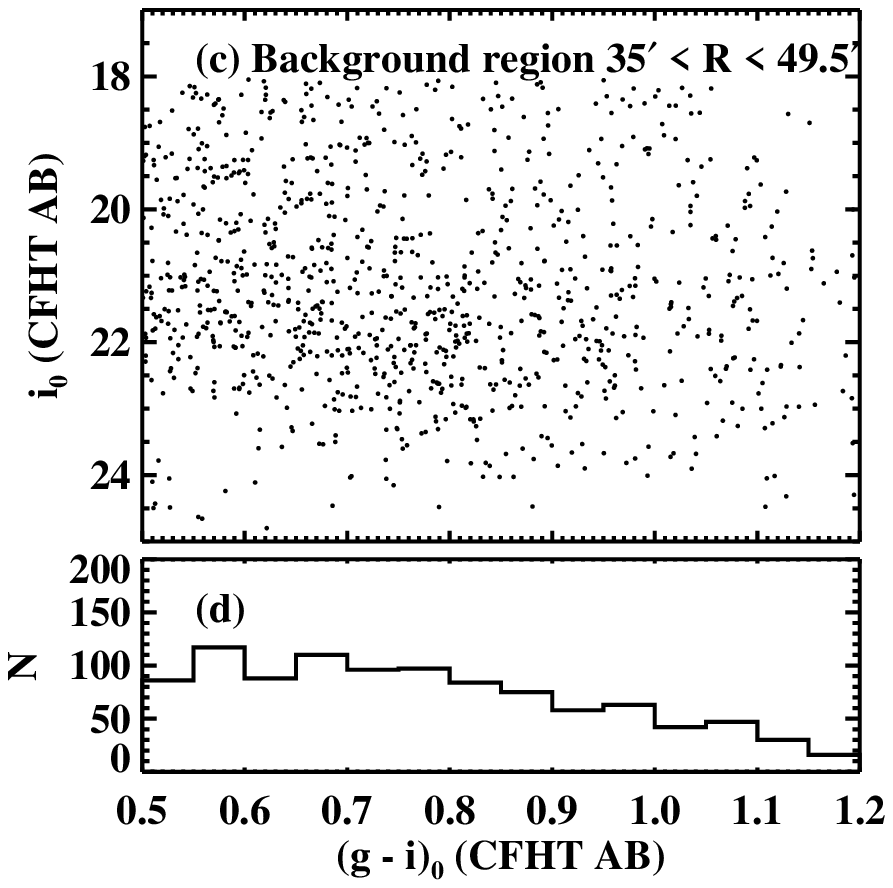}
	\includegraphics[scale=0.65]{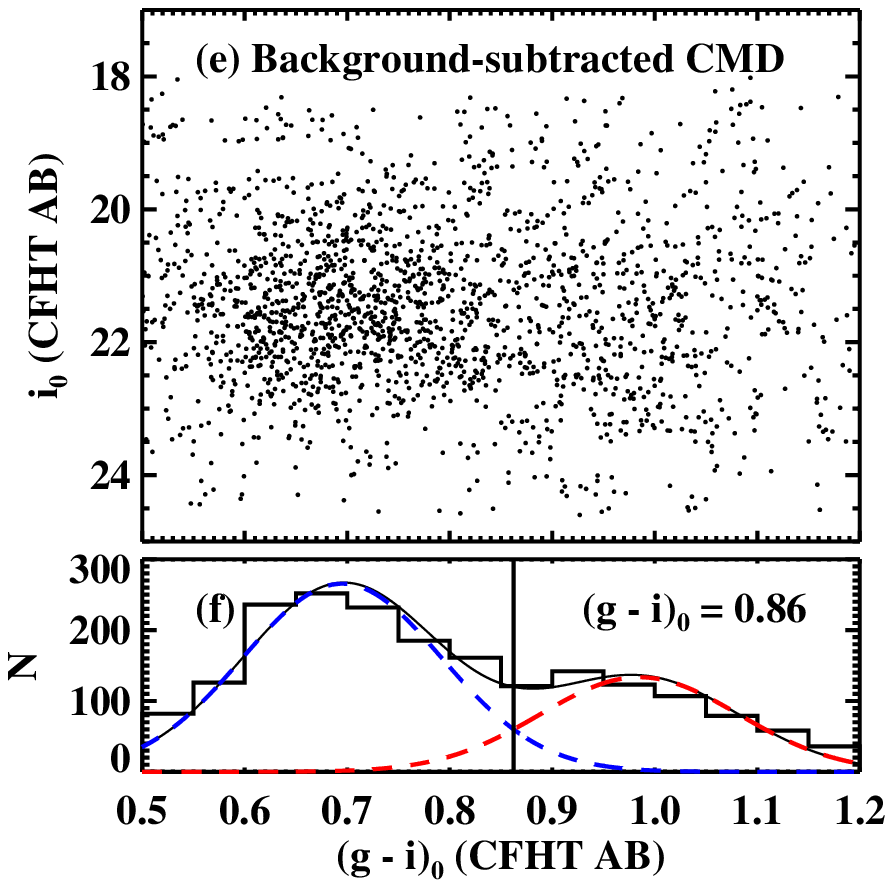}
	\includegraphics[scale=0.65]{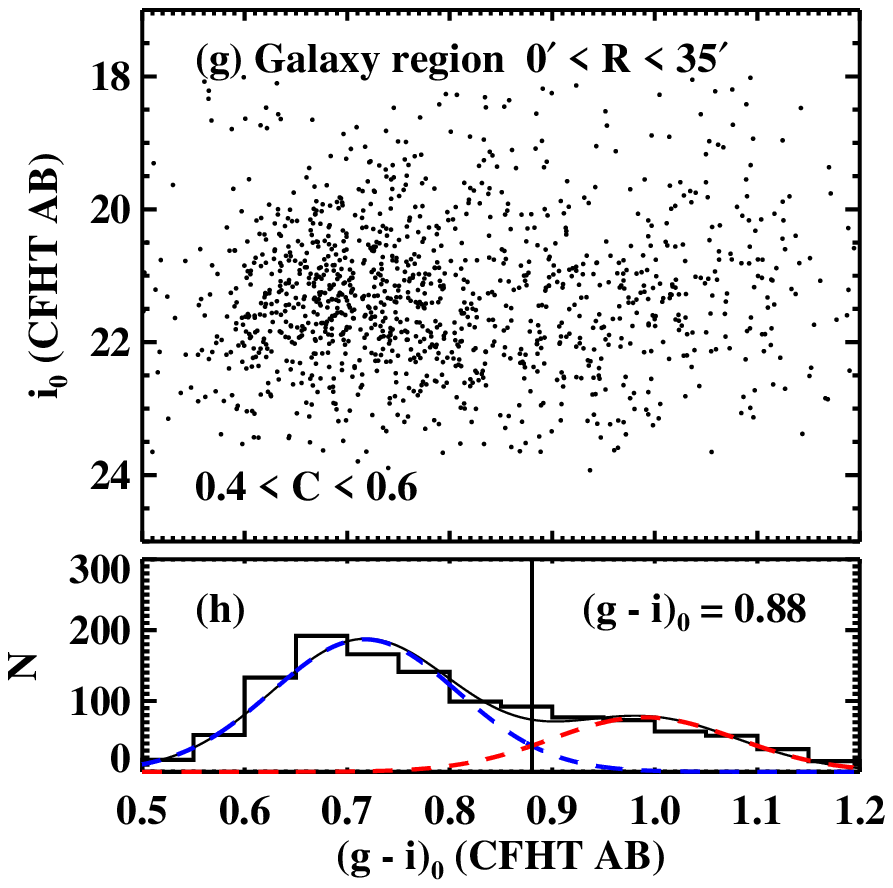}
	\caption{
		(a,b) $i_0-(g-i)_0$ CMD and $(g-i)_0$ color distribution of the GC candidates in the galaxy region ($0'<R<35'$).
		(c,d) Same as left but for the GC candidates in the background region ($35'<R<49\farcm5$).
		(e,f) Background-subtracted CMD and color distribution of the GC candidates. 
		We also plot the gaussian fits derived from the GMM test on the histogram. 
		According to the GMM results, blue and red subpopulations are divided at $(g-i)_0=0.86$. 
		(g,h) Same as (a,b) but for the compact GC candidates with $0.4<C<0.6$. 
	}
	\label{fig:cmd}
\end{figure*}

Finally, we set the galactocentric distance limit to select our final GC candidates. 
Figure \ref{fig:rdp0} shows the radial number density profile of the GC candidates along the galactocentric distance. 
Here, the GC candidates are the sources within the dashed trapezoid marked in Figure \ref{fig:cindex2}(a). 
As shown in Figure \ref{fig:comp}(b), the central region of the galaxy at $R<4'$ is incomplete to find the GC candidates 
due to the bright galaxy light in our data,
so we use HST/ACS data to supplement the number density of the central region at $R<4'$.
After constructing the complete profile from the center to the outer region, we fit the profile with the S\'ersic function including the background. 
We obtain the effective radius of the GC system of $R_{e,GCS}=7\farcm59\pm0.72$, S\'ersic index of $n=2.70\pm0.21$, and the background level of $0.39\pm0.01$ sources/arcmin$^2$.
This size of the GC system is larger than the 
size derived from previous studies \citep[$R_{e,GCS}\approx6'$,][]{har14}. 
We set the radial coverage limit to about $5R_{e,GCS}\approx35'$ where the profile 
begins to 
flatten. Moreover, our follow-up study shows that the spectroscopically confirmed GCs are mostly located out to $R=35'$ (Kang et al. 2022, in prep.).

\subsection{Color-magnitude Diagram of GC Candidates}

Figure \ref{fig:cmd}(a) shows the $i_0-(g-i)_0$ color-magnitude diagram (CMD) of the GC candidates in the galaxy region at $0'<R<35'$. 
Because we include point sources when selecting GC candidates, there should be foreground stars in the CMD. 
To statistically remove any foreground or background contamination, we check the CMD of the GC candidates in the background region at $35'<R<49\farcm5$, as shown in Figure \ref{fig:cmd}(c). 
The boundary for the background region, $R=49\farcm5$, is determined so that the galaxy region and the background region have the same area. 
Then we statistically subtract the CMD 
of the background region from the CMD 
of the galaxy region to obtain our final CMD for the GC candidates. 
We divide the CMD into small grids with a size of $\Delta i_0=0.5$ and $\Delta (g-i)_0=0.05$, and we subtract the number of background sources from the number of sources in the galaxy region. 
The size of the grid or the selection of the subtracted sources does not change the results much.
In Figure \ref{fig:cmd}(e), we show the CMD after background subtraction. 
Note that the blue GC population is dominant in the color range of $0.5<(g-i)_0\lesssim0.9$ and the red GC population is located at $0.9\lesssim(g-i)_0<1.2$.

We compare the results of two methods of selecting GC candidates: (1) selecting point and compact sources ($0.3<C<0.6$) and then subtracting the contaminants, and (2) selecting only compact sources ($0.4<C<0.6$).
In Figure \ref{fig:cmd}(g) we show the CMD of the compact GC candidates in the galaxy region.
This is very similar to Figure \ref{fig:cmd}(e), meaning that the compact GC candidates are mostly genuine GCs, and the two methods are not much different. 

\begin{deluxetable*}{cccccccccc}[htb!]
	\tabletypesize{\scriptsize}
	\tablecaption{A Summary of GMM Tests for Color Distributions of the GCs in M104 \label{table:gmm}}
	\tablewidth{0pt}
	\tablehead{
		\colhead{Option} & \colhead{Color} & \colhead{$\mu_{blue}$} & \colhead{$\sigma_{blue}$} & \colhead{$N_{tot,blue}$} & \colhead{$\mu_{red}$} & \colhead{$\sigma_{red}$} & \colhead{$N_{tot,red}$} & \colhead{$p$} & \colhead{$D$}
	}
	\startdata
	$\sigma_{blue}=\sigma_{red}$ & $(g-i)_0$ & $0.696\pm 0.003$ & $0.097\pm 0.002$ & $1291\pm 28$ & $0.985\pm 0.006$ & $0.097\pm 0.002$ & $649\pm 28$ & $<0.01$ & $2.98\pm 0.09$ \\
	$\sigma_{blue}\ne\sigma_{red}$ & $(g-i)_0$ & $0.681\pm 0.007$ & $0.087\pm 0.004$ & $1158\pm 70$ & $0.957\pm 0.015$ & $0.112\pm 0.007$ & $782\pm 70$ & $<0.01$ & $2.75\pm 0.17$ \\
	\hline
	\enddata
\end{deluxetable*}


\subsection{Color Distribution of GC Candidates}

The lower panels of Figure \ref{fig:cmd} show a $(g-i)_0$ 
color histogram of each CMD.
Note that the GC candidates in the galaxy region, the GC candidates after subtracting the contamination, 
and the compact GC candidates in the galaxy region all 
show a bimodal distribution. 
To test and quantify the color bimodality, we perform Gaussian Mixture Modeling (GMM) test 
with the same variance option 
\citep{mur10} to the background-subtracted GC candidates. 
As a result, the unimodal distribution was rejected ($p$-value $p<0.01$, separation $D=2.98 \pm 0.09$),   
meaning that the data shows statistically meaningful bimodal distributions.
The results for the GMM test are summarized in Table \ref{table:gmm}. 
The two Gaussian functions with peaks at $(g-i)_0 = 0.696\pm0.003$ and $(g-i)_0 = 0.985\pm0.006$ are overplotted to the histogram, 
and we can divide the blue GCs and the red GCs using $(g-i)_0=0.86$. 
We also perform the GMM test with the different variance option, and we summarized the results in Table \ref{table:gmm}.
If we choose the different variance option, the Gaussian width for the red GCs is abnormally large. 
Therefore, we adopt the same variance option for the following analysis.

\begin{figure}[tb!]
    \centering
	\includegraphics[scale=0.9]{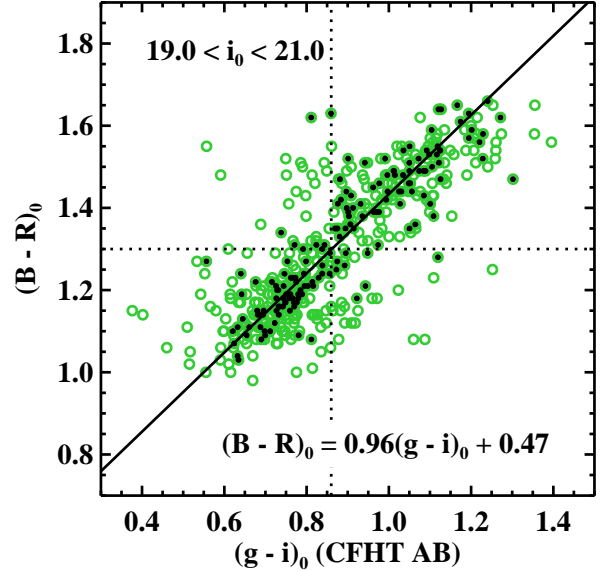}
	\caption{
	    $(B-R)_0$ vs. $(g-i)_0$ color-color relation of the GCs matched with \citet{har10}. 
	    Green circles mark all the GCs matched with \citet{har10} and black filled circles mark the bright GCs with $19<i_0<21$ mag.
	    To obtain the relation between the two colors, we fit the linear function for the bright GCs and plot the function as a solid line.
	    Dotted lines mark the color dividing blue and red subpopulations, $(g-i)_0=0.86$ and $(B-R)_0=1.3$. 
	}
	\label{fig:ccd2}
\end{figure}

If we assume that the ages of the GCs are as old as Milky Way GCs, 
then color bimodality indicates the metallicity bimodality and shows blue metal-poor GCs and red metal-rich GCs.
To derive the linear relation between the color and the metallicity, 
we follow the relation of \citet{spi06} derived from Milky Way GCs, [Fe/H] = $3.06(B-R)_0 - 4.90$.
In Figure \ref{fig:ccd2}, we show $(B-R)_0$ vs. $(g-i)_0$ color of the sources matched with \citet{har10} and the linear relation derived from the bright sources with $19<i_0<21$, $(B-R)_0 = 0.96(g-i)_0+0.47$ (RMS=0.08). 
In this relation, $(g-i)_0=0.86$ corresponds to $(B-R)_0=1.3$. 
These colors are used to divide the two subpopulations in this study and \citet{spi06,har10}, respectively, meaning that our study is very consistent with these previous studies. 
Using this relation, we transform $(B-R)_0$ color to $(g-i)_0$ color to get the following relation: 
[Fe/H] = $2.94(g-i)_0 - 3.46$. 
As a result, the metallicity range of the GC candidates is about $-2.0 <$ [Fe/H] $< +0.1$ 
and the mean metallicities for the blue and red subpopulations correspond to [Fe/H] = $-1.4$ and $-0.6$. 
These results are consistent with the results for the Milky Way GCs \citep{har96}.



\begin{figure}[htb!]
    \centering
	\includegraphics[scale=0.9]{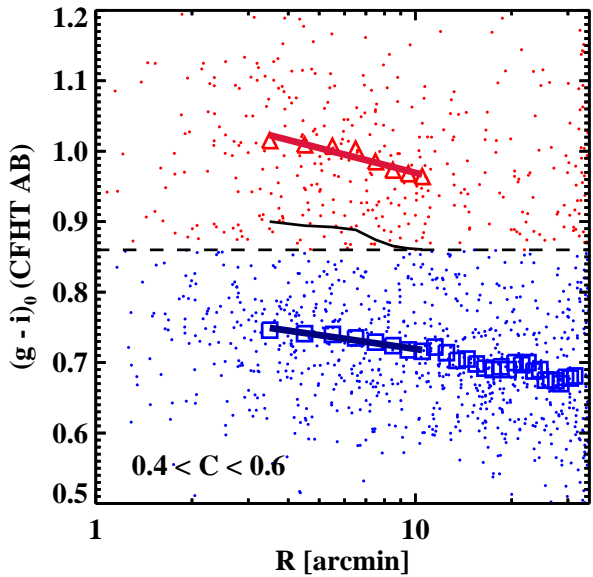}
	\caption{
	    $(g-i)_0$ color distribution of the compact GC candidates ($0.4<C<0.6$) along the galactocentric distance. Blue squares and red triangles mark the Gaussian peaks for each radial bin derived from the GMM results. 
	    Blue and red solid lines mark the color gradient. 
	    Black solid line marks the color dividing blue and red subpopulation for each radial bin.
	    We adopt $(g-i)_0=0.86$ marked by the dashed line to divide blue and red subpopulations for all radial range. 
	}
	\label{fig:coldist}	
\end{figure}

Figure \ref{fig:coldist} shows the $(g-i)_0$ color distributions of the GC candidates along the galactocentric radius. 
Here we only use the clean samples of compact GC candidates with $0.4<C<0.6$. 
In each radial bin from $0'<R<7'$ to $28'<R<35'$ with one arcminute steps, 
we mark the blue/red peak values of the Gaussian models derived from the GMM test. 
Note that a color gradient is seen for both the blue and the red GC subpopulations. 
The blue GC subpopulation is broadly located, and we measure the color gradient with linear fitting, $\Delta(g-i)_0/\Delta({\rm log}R)=-0.07\pm0.01$ mag dex$^{-1}$ at $R<10'$. 
The red GC subpopulation is mostly located at $R<10'$ and the color gradient in that region is as large as $\Delta(g-i)_0/\Delta({\rm log}R)=-0.12\pm0.01$ mag dex$^{-1}$.
\citet{har14} found the color gradient of $\Delta(B-R)_0/\Delta({\rm log}R)=-0.06\pm0.01$ mag dex$^{-1}$ and $-0.05\pm0.01$ mag dex$^{-1}$ for the blue and red subpopulations within $R<7'$. 

Compared to the results of \citet{har14}, our results for the blue subpopulation 
show a similar degree of color gradient. 
For the red subpopulation, 
the degree of gradient is larger in this study. 
This difference can be caused by the difference 
in foreground contamination control. 
We measure the red peak after 
removing point sources but the previous study did not. 



\subsection{Radial Distributions of the GC Subpopulations}

\begin{figure*}[htb!]
    \centering
	\includegraphics[scale=1.0]{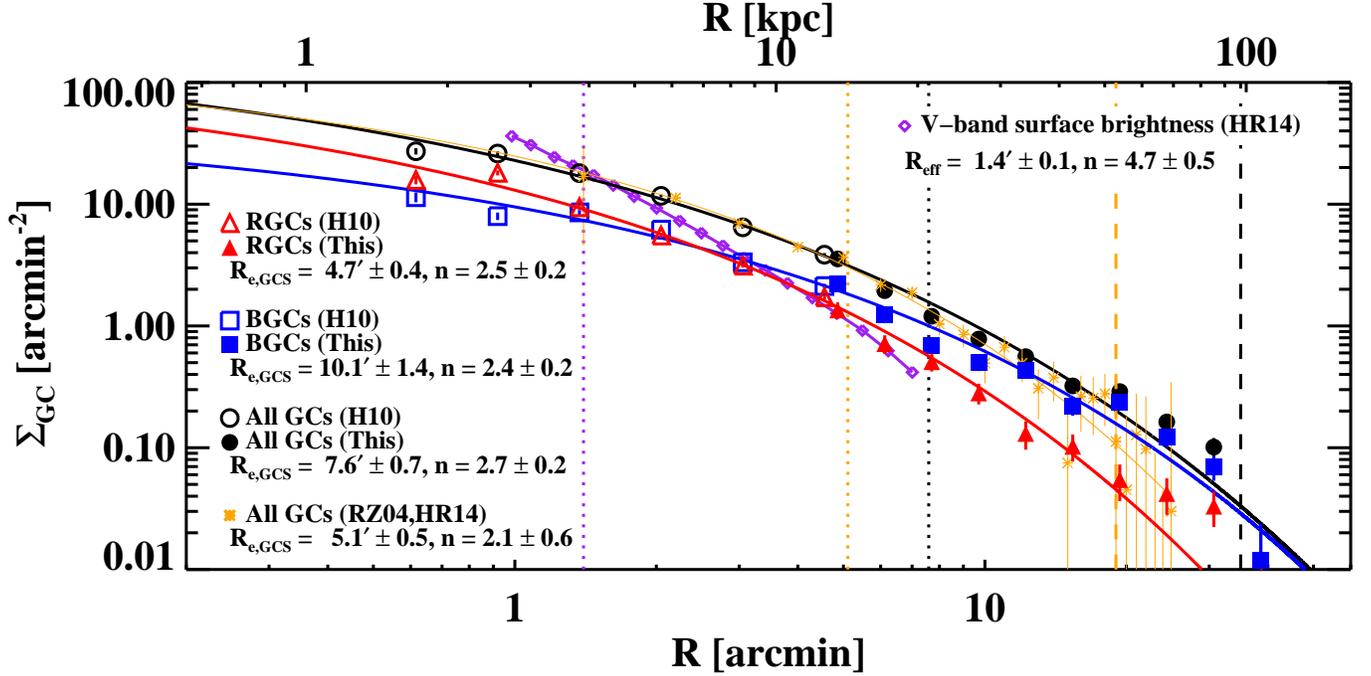}
	\caption{
	    Background-subtracted radial number density profile of GC candidates 
	    according to galactocentric distance.
	    GC subpopulations are divided at $(g-i)_0 = 0.86$. 
	    As in Figure \ref{fig:rdp0}, we use the GCs found in \citet{har10} (open symbols) to supplement the GC candidates from this study (filled symbols). 
        Black circles mark the entire GC candidates, and red triangles and blue squares mark red and blue GC candidates, respectively.
        Solid lines mark the fitted S\'ersic profile, vertical dotted lines mark the effective radius of the GC system, and vertical dashed lines mark the radial coverage limit of the GCs. 
        Yellow asterisk symbols mark the profile of the GC candidates from previous studies \citep{rho04,har14} shifted to be consistent with our results.
        Violet diamonds mark the V-band galaxy light surface brightness \citep{har14} shifted to be consistent with our results. 
	}
	\label{fig:rdp}	
\end{figure*}

Based on the result shown in Figure \ref{fig:cmd}(e,f), we divide the entire 
selection of GC candidates into two subpopulations, 
blue GCs ($0.50 \leq (g-i)_0 < 0.86$) and red GCs ($0.86 \leq (g-i)_0 < 1.20$),
and we examine their radial distributions separately. 
Figure \ref{fig:rdp} shows the background-subtracted radial number density profile of the GC candidates along the galactocentric distance. 
We fit the profiles with S\'ersic function 
$\Sigma_{GC}(R)=N_e~{\rm exp}[-b_n[(R/R_{e,GCS})^{1/n}-1]]+bg$ and then subtract the background level. 
The fitting results are summarized in Table \ref{table:rdp}.
The S\'ersic index is $n=2-3$ which is in agreement with the previous study. 
We plot 
the GC radial profile and the $V$-band galaxy light surface brightness profile from \citet{har14}, 
where the zeropoint is 
arbitrary shifted to overlap 
with the number density 
of the GC candidates at $R\approx 5'$.

Several 
features are noticeable. 
First, the profile of all GCs 
at $R < 20'$ 
agrees well with the profile from \citet{har14},
but our profile probes 
 further 
than the latter. 
Second, 
the number density of the red GCs is larger than that of the blue GCs 
in the central region at $R<3'$. 
Third, the number density of the red GCs decreases 
 more quickly than that of the blue GCs, 
which means that the red GCs are more centrally concentrated.
This result corresponds to previous studies \citep[Figures 23 and 24 of][]{har14}.
Fourth, the profile of the galaxy light follows the profile of the red GCs better than the profile of the blue GCs at $R\gtrsim3'$, but follows neither at $R\lesssim3'$.
Lastly, there exists a break at $R\approx20'$  
and the GC profiles of the outer region at $R \approx 20'-35'$ show a slight excess of GCs over the fitted lines. This excess is mainly due to blue GCs.
Based on these results, we can divide the entire region into three regions:  
(a) $R\lesssim3'$ where the stellar light is dominant, and red GCs are more dominant than the blue GCs,
(b) $3'\lesssim R \lesssim20'$ where the stellar light follows the red GCs, and the blue GCs 
become more dominant, and
(c) $20'\lesssim R \lesssim35'$ where the blue GCs are more dominant.

\begin{deluxetable}{ccccc}[b!]
	\tabletypesize{\scriptsize}
	\tablecaption{A Summary of S\'ersic Fits for the Radial Number Density Profiles of the GCs in M104	\label{table:rdp}}
	\tablewidth{0pt}
	\tablehead{
		\colhead{Sample} & \colhead{$N_e$} & \colhead{$n$} & \colhead{$R_{e,GCS}$} & 
		\colhead{bg} \\
    	\colhead{} & \colhead{(arcmin$^{-1}$)} & \colhead{} & \colhead{(arcmin)} & 
    	\colhead{(arcmin$^{-1}$)}
	}
	\startdata
	All GCs & $1.57\pm 0.24$ & $2.70\pm 0.21$ & $7.59\pm 0.72$ & 
	$0.39\pm 0.01$ \\
	Blue GCs & $0.61\pm 0.13$ & $2.37\pm 0.24$ & $10.08\pm 1.37$ & 
	$0.26\pm 0.01$ \\
	Red GCs & $1.53\pm 0.22$ & $2.47\pm 0.23$ & $4.68\pm 0.38$ & 
	$0.13\pm 0.01$ \\
	\hline
	\enddata
\end{deluxetable}

From the effective radius of the GC system derived from the S\'ersic fit, we can estimate the stellar mass of M104. 
Using 
the relation between the effective radius of a GC system and the stellar mass of its host galaxy in \citet{for18}, log$(R_{e,GCS})$ [kpc] $= 0.97$ log$(M_{*}/M_\odot) -9.76$, 
we obtain $M_{*}\simeq 2.7 \times 10^{11} M_\odot$. 
This value is consistent with the previous estimations: $M_*=1.79\times10^{11}M_\odot$ based on \textit{Spitzer} IRAC 3.6$\mu m$ imaging \citet{mun15} and
$M_*=2.1\times10^{11}M_\odot$ based on $K$-band magnitude \citep{kar20}.

\subsection{Spatial Distributions of the GC Subpopulations}

\begin{figure*}[htb!]
    \centering
	\includegraphics[scale=1.0]{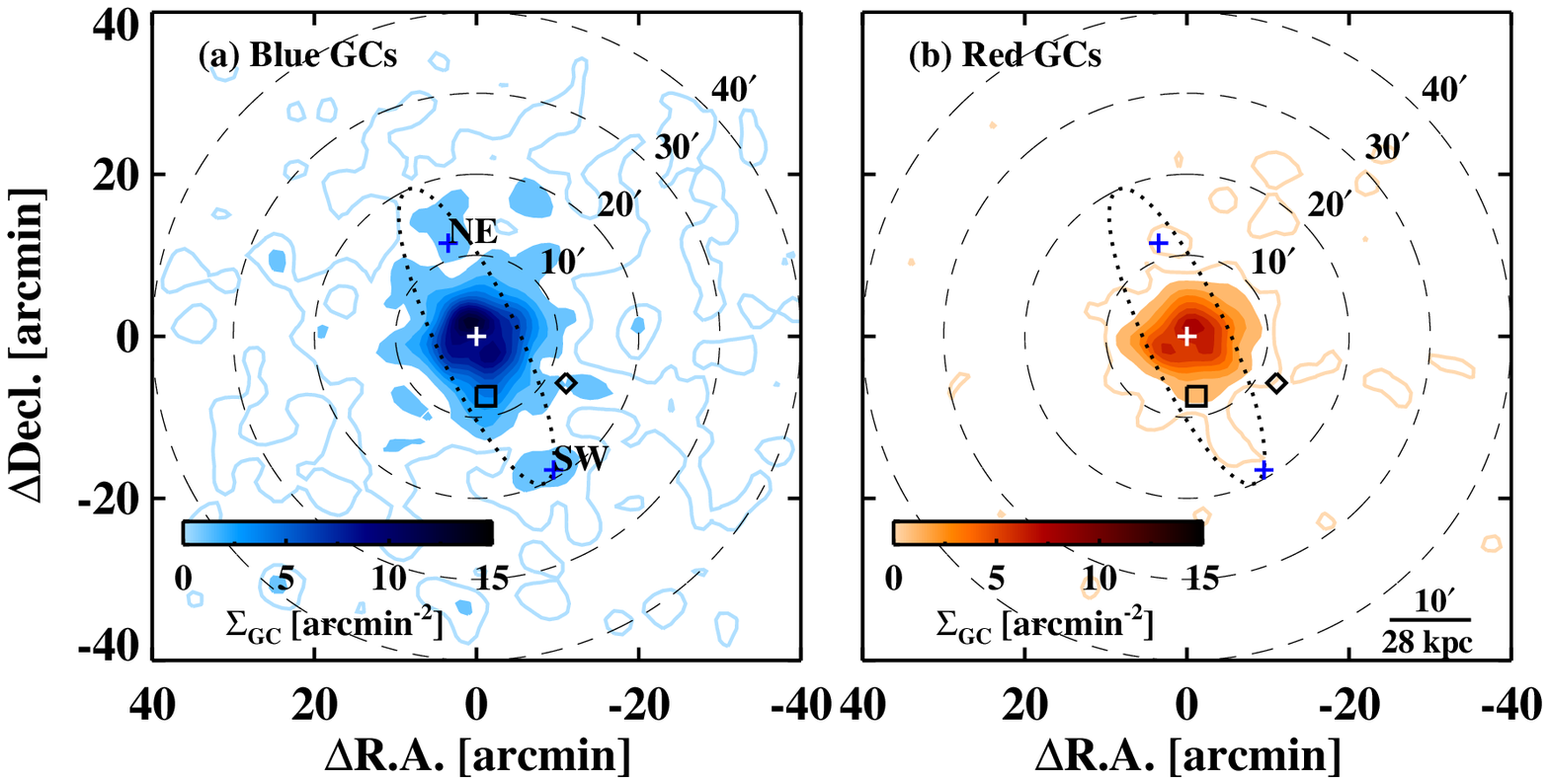}
	\includegraphics[scale=1.0]{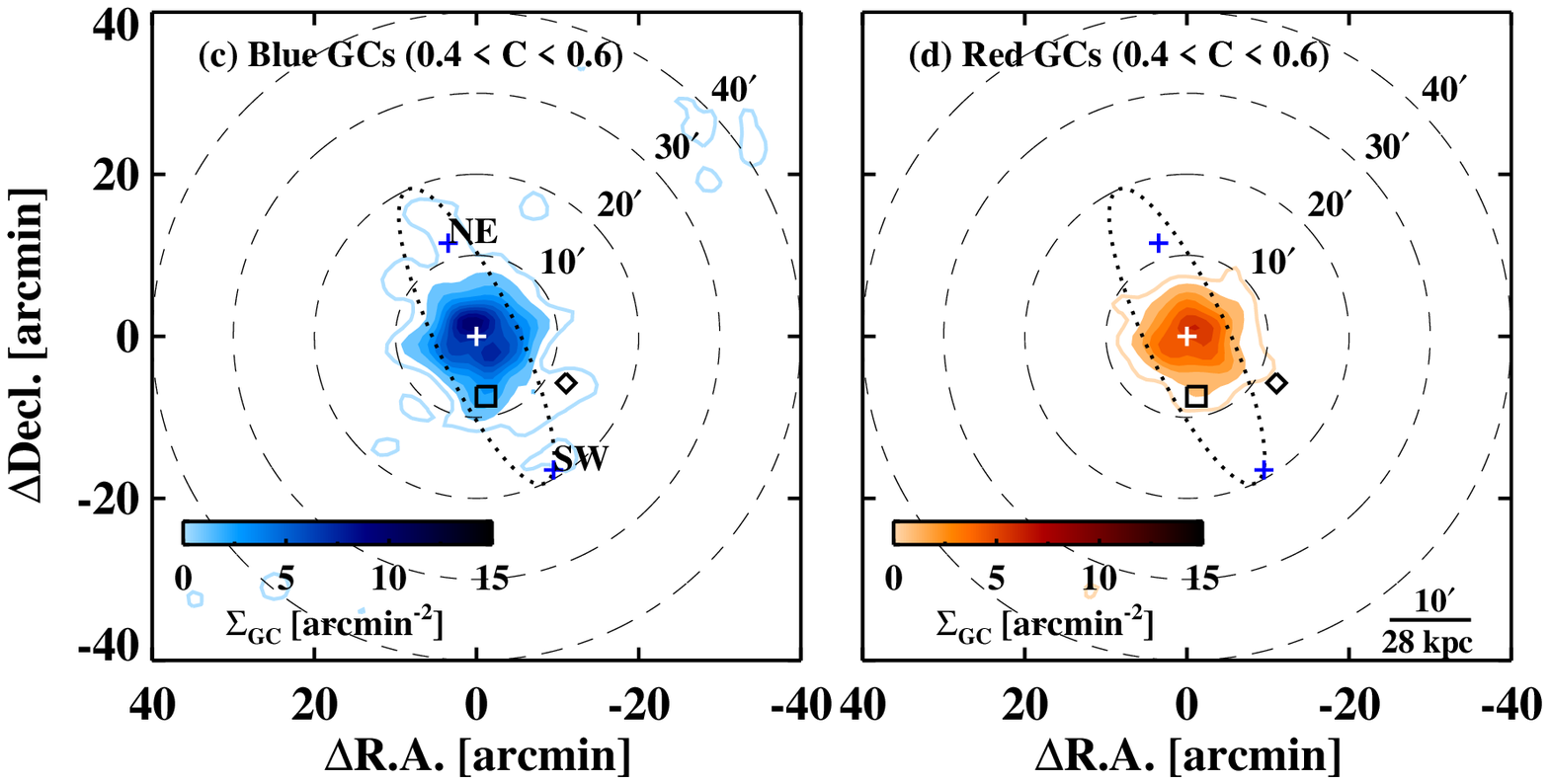}
	\caption{
		Surface number density maps of (a) blue GC candidates and (b) red GC candidates. 
		Dashed circles mark the circles with 10$\arcmin$, 20$\arcmin$, and 30$\arcmin$ galactocentric radii.
		White plus symbols mark the galaxy center and blue plus symbols mark the center of the blue GC substructures. 
		Dotted ellipses mark the stream (loop) schematically. 
        Squares and diamonds mark the position of the dwarf galaxies NGC 4594-DGSAT1 and NGC 4594 DW1.
    } 
	\label{fig:spatial}	
\end{figure*}

Figure \ref{fig:spatial} shows the spatial distribution of each subpopulation. 
The open contour maps indicate 
the 1 arcmin$^{-2}$ level and the filled contour maps indicate 
the 2 to 15 arcmin$^{-2}$ levels. 
First, 
the distributions of the GC candidates in the central region 
are almost circular for both the blue and the red systems.
To check if there are any substructures in the central region, we use the ACS data to derive the spatial distributions of the GCs, and we find no significant substructures. 
Second, the blue GCs are much more widely distributed than the red GCs, out to $R>30'$. 
This result is consistent with the previous findings of massive ETGs, and supports the external origin 
scenario of the blue GCs. 
Third, 
there are two major blue GC substructures at the north-east and the south-west direction (called NE BGC clump and SW BGC clump, respectively). 
The locations of these two substructures are very close to the faint substructures reported in \citet{mal97a} and \citet{mal97b}. 
The recent paper of \citet{mar21} revealed the full path of the faint stream for the first time. 
We schematically mark this stream with a dotted ellipse, and we confirm that the location of the blue GC clumps at $10'<R<20'$ is well matched with the location of the loop edges. 
Further analysis of the blue GC clumps will be described in Section \ref{sec:substructure}.

We also check the spatial distribution of each subpopulation for the compact GC candidates. 
Due to the small number, most features that are widely distributed out to $R>30'$ have disappeared.
However, the overall features remain, including the two substructures. 

\subsection{Luminosity Function of GCs} \label{sec:gclf}

\begin{figure*}[htb!]
    \centering
	\includegraphics[scale=0.65]{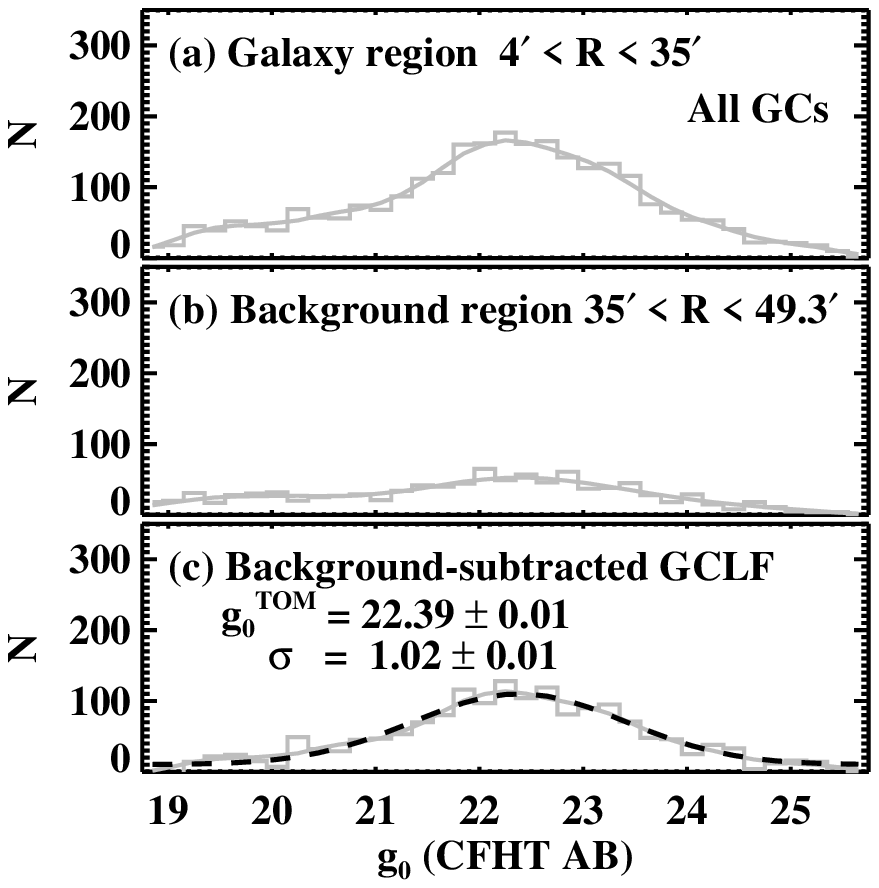}
	\includegraphics[scale=0.65]{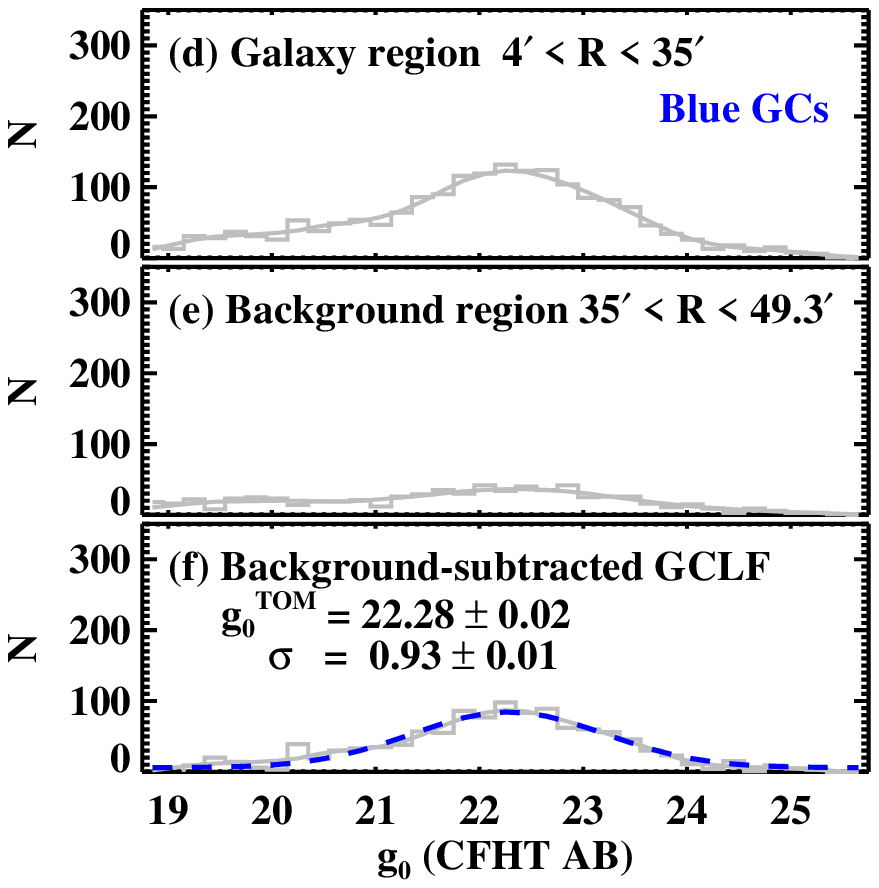}
	\includegraphics[scale=0.65]{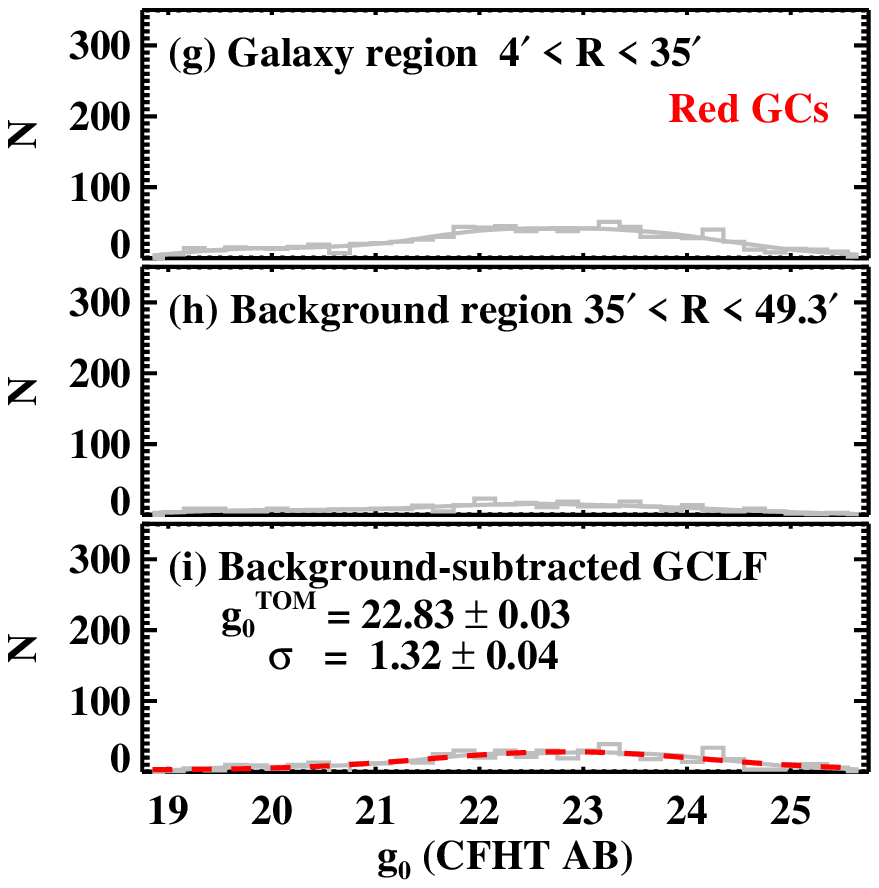}
	\includegraphics[scale=0.65]{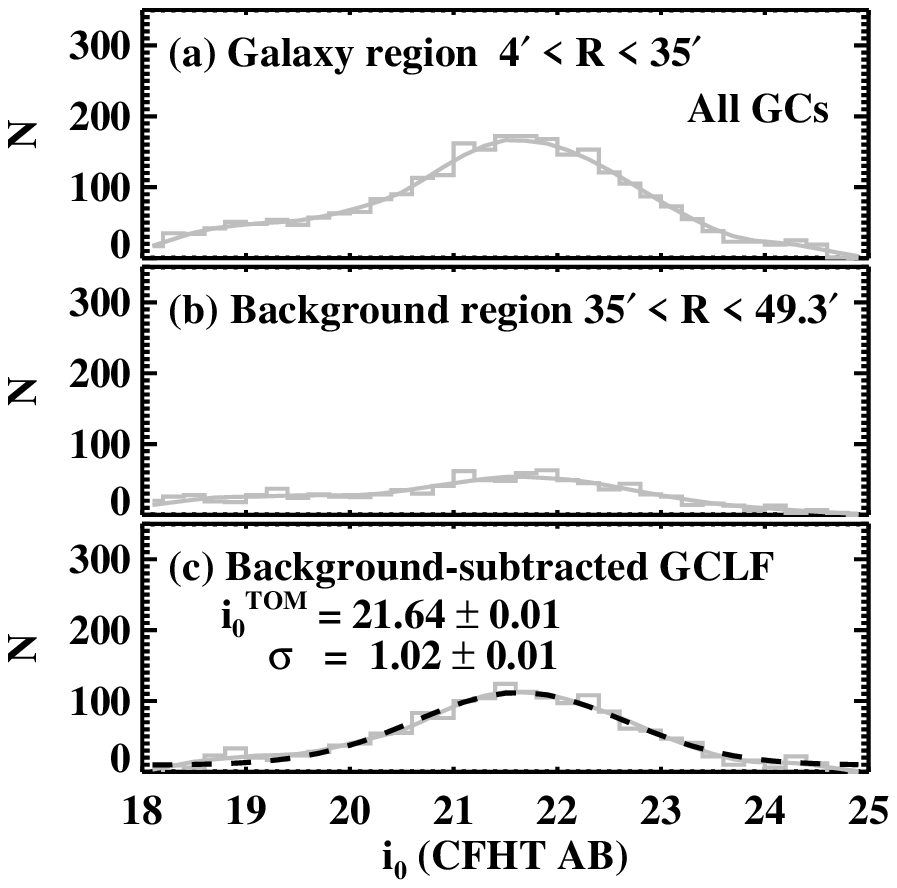}
	\includegraphics[scale=0.65]{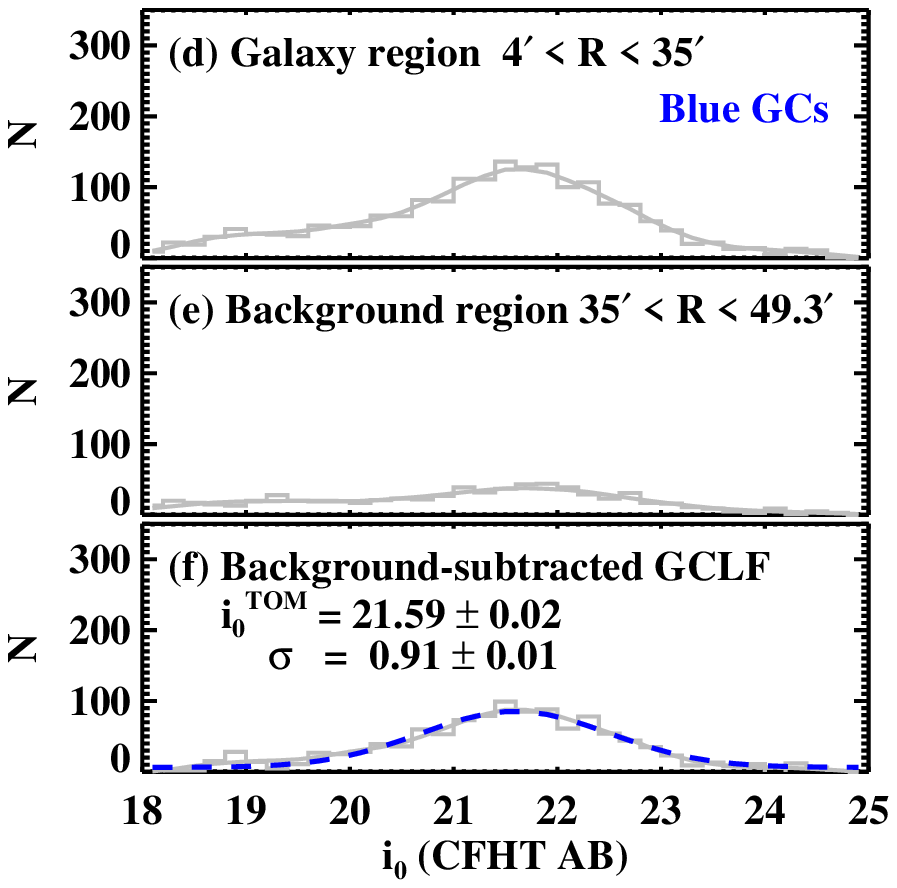}
	\includegraphics[scale=0.65]{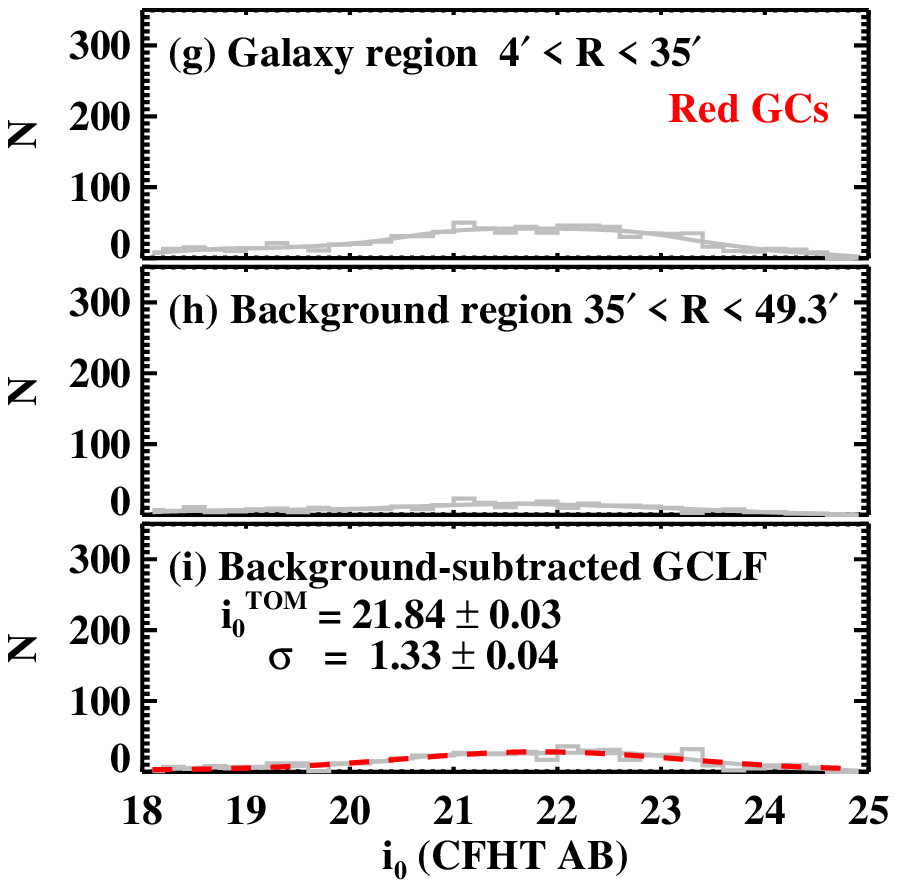}
	\caption{
		$g$ and $i$-band luminosity function of the GC candidates in M104.
		Galaxy region GCLF, background region GCLF, background-subtracted GCLF are shown.
		Grey lines mark the KDE results and histograms.
		Dashed lines mark the Gaussian fitting results. 
	}
	\label{fig:gclf}
\end{figure*}

We derive the GCLF from the photometry of the GC candidates identified in this study. 
The left panels of Figure \ref{fig:gclf} show 
 the $g$ and $i$-band GCLFs of the galaxy region and the background region separately. 
We exclude the innermost incomplete region at $R<4\arcmin$ from the galaxy region and only use the region at $R>4\arcmin$. 
We set the background region to have the same area as the galaxy region and subtract the background region LF from the galaxy region LF to obtain the background-subtracted GCLF. 
Finally we derive TOMs with two methods, Gaussian fit and kernel density estimation, and find no difference: 
$g_0^{\rm TOM} = 22.39\pm0.01$ and  
$i_0^{\rm TOM} = 21.64\pm0.01$
with $\sigma=1.02\pm0.01$. 
We summarize the Gaussian fitting results for the GCLFs in Table \ref{table:gclf}.

\begin{deluxetable*}{cccrccr}[htb!]
	\tabletypesize{\scriptsize}
	\tablecaption{A Summary of Gaussian Fits for GCLFs in M104	\label{table:gclf}}
	\tablewidth{0pt}
	\tablehead{
	    \colhead{} & \colhead{$g_0^{TOM}$} & \colhead{$\sigma_g$} & \colhead{$N_{tot,g}$} & \colhead{$i_0^{TOM}$} & \colhead{$\sigma_i$} & \colhead{$N_{tot,i}$}
	}
	\startdata
	All GCs & $22.39\pm 0.01$ & $1.02\pm 0.01$ & $1640\pm 10$ & $21.64\pm 0.01$ & $1.02\pm 0.01$ & $1640\pm 10$ \\
	Blue GCs & $22.28\pm 0.02$ & $0.93\pm 0.01$ & $1120\pm 10$ & $21.59\pm 0.02$ & $0.91\pm 0.01$ & $1120\pm 10$ \\
	Red GCs & $22.83\pm 0.03$ & $1.32\pm 0.04$ & $520\pm 10$ & $21.84\pm 0.03$ & $1.33\pm 0.05$ & $520\pm 10$ \\
	\enddata
\end{deluxetable*}

Based on the 
deep photometry of the ACS field centered on M104, \citet{spi06} derived TOM of the M104 GCLF:  
$V_0^{\rm TOM}=22.17\pm0.06$ from the kernel density estimate and $22.03\pm0.06$ from the Gaussian estimate. 
This corresponds to $M_V^{\rm TOM} = -7.60\pm0.06$ and $-7.74\pm0.06$ for their adopted distance modulus of $(m-M)_0=29.77\pm0.03$, respectively. 
These values will be $M_V^{\rm TOM} = -7.73\pm0.10$ and $-7.87\pm0.10$ for the TRGB distance of $(m-M)_0=29.90\pm0.08$ adopted in this study \citep{mcq16}.
They are consistent with the GCLF calibration based on the Milky Way GCs \citep{dic06,rej12,lee19}.
Moreover, we successfully reproduce the GCLF results of \citet{spi06} using the catalog provided by them. 
If we convert 
\citet{spi06}'s results based on $V$-band magnitudes to $g$-band magnitudes using the relation 
$g=V+0.23(B-R)$ derived from the bright GCs matched with \citet{spi06}, we obtain $g_0^{\rm TOM} = 22.40\pm0.01$ which is consistent with our results.

We also show the GCLFs for the blue and red subpopulations in the middle and the right panels of Figure \ref{fig:gclf}. 
The fitting results are summarized in Table \ref{table:gclf}. 
From the results, we find significant TOM differences between the blue and red subpopulations in both filters, 
$\Delta g_0^{\rm TOM}\rm (BGC-RGC)=-0.55\pm0.04$ mag and 
$\Delta i_0^{\rm TOM}\rm (BGC-RGC)=-0.25\pm0.04$ mag.
This trend that 
the blue GCs are on average brighter than red GCs can be explained by the metallicity difference assuming the same mass \citep{ash95}, and is shown for nearby ETGs \citep{lar01a}. 
However, \citet{spi06} found that the TOMs of the blue GCs and the red GCs in M104 are not much different with
$\Delta V_0^{\rm TOM}\rm (BGC-RGC)=-0.09\pm0.12$ mag. 
If we only select the GCs from the CFHT data within the ACS coverage, we do not see the trend anymore. This means that the GCs outside the ACS coverage made the TOM difference. It is hard to say that this difference is due to any contamination effect of wide-field photometry or whether it is physically meaningful. 
Moreover, recent studies of the GCLFs for NGC 4921, NGC 4874, and NGC 4889 \citep{lee16} and NGC 4589 \citep{lee19} also showed no difference between the two populations. 
GCLF studies for more ETGs with wide-field imaging and spectroscopy will be helpful to resolve this issue. 



\subsection{Total Number and Specific Frequency of GCs}
By integrating the radial number density profile of the GCs with $i_0<i_0^{TOM}=21.64$ to $R<35'$ and 
 doubling the number, we obtain the total number of GCs in M104:  
$N_{GC}=1610\pm30$ and $S_N=1.8\pm0.1$. 
\citet{rho04} presented  slightly larger values but with larger errors: 
$N_{GC}=1900\pm200$ and $S_N=2.1\pm0.3$.
Although the integration range of this study is larger than that of \citet{rho04}, we obtain a smaller value. 
This difference may be due to the different color selection criteria between the two studies. 
Note that we use $ugi$ bands for GC selection, while \citet{rho04} used $BVR$ bands.
The total number of GCs in NGC 5128 is $N_{GC}=1450\pm160$ \citep{hug21} which is very similar to that of M104. 

The total number of GCs in each subpopulation is $N_{BGC}=1060\pm20$ and $N_{RGC}=550\pm20$, so we obtain the fraction of red GCs of M104 to be $f_{RGC}=0.34\pm0.01$. 
This value is smaller than the values derived from previous studies \citep[$f_{RGC}=0.40$ to 0.55,][]{lar01b}. 
because they did not cover the outer region 
where blue GCs are more abundant. 
The fraction of red GCs in M104 is similar to that of the ETGs in the Virgo with similar mass or luminosity to M104
\citep{pen08}.


\section{Discussion} \label{sec:discuss}









\subsection{Dual Halos of M104}

Noting a significant difference in spatial structures (especially the ellipticity) between the blue and red GC systems of Virgo ETGs based on homegeneous HST/ACS data, \citet{par13} argued that massive ETGs have dual halos, a metal-rich halo embedded in an outer metal-poor halo.  
\citet{par13} also argued that dual halos formed in different merging processes. 
The metal-rich halo forms via wet merging or dry merging of relatively massive galaxies 
 and the metal-poor halo grows via accretion of satellite dwarf galaxies.

From this study, 
we 
find that there are also dual halos in M104.
The outer halo of M104 expected from the previous studies of stellar light is small ($R\sim9'$) and red (indicating metal-rich).
In the radial density profile as seen in Figure \ref{fig:rdp}, the inner region at $3'\lesssim R \lesssim20'$ where the stellar light follows the red GCs 
corresponds to an inner metal-rich halo. So, the outer halo of M104 expected in the previous studies is consistent with this metal-rich halo.
The outer region at $20'\lesssim R \lesssim35'$ where the blue GCs are dominant 
corresponds to an outer metal-poor halo.
This giant outer structure was hard to find in the previous studies due to their narrow spatial coverage. 
These two halos in M104 might have formed  according to  the formation scenarios as described in \citet{par13}.

\subsection{Blue GC Clumps and the Progenitor of the Stellar Stream around M104} \label{sec:substructure}

\begin{figure}[b!]
    \centering
	\includegraphics[scale=0.9]{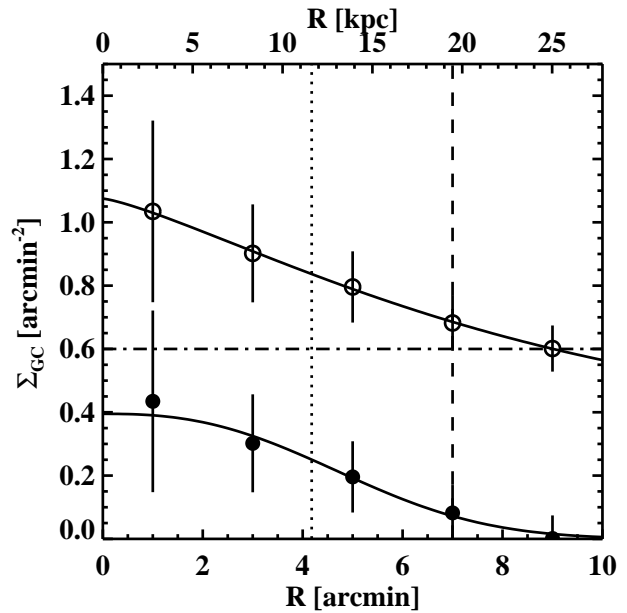}
	\caption{Radial number density profile of the GC candidates around the south-west clump before background subtraction (open symbols) and after background subtraction (filled symbols).
	Solid lines mark the S\'ersic fittings, dot-dashed line marks the background level, dashed line marks the boundary for the substructure, and dotted line marks the effective radius of the substructure.
	}
	\label{fig:bgcsub}	
\end{figure}

Here we analyze the blue GC clumps,
especially 
the SW BGC clump
found near the south-western stellar stream. 
\citet{mar21} interpreted the existence of the SW stream by a single disruption event 
of a dwarf progenitor ($\sim10^8 M_\odot$) about 3 Gyr ago. 
If so, the SW BGC clump 
may be the remnant of the dwarf progenitor. 
In Figure \ref{fig:bgcsub} 
we plot the radial density profile 
of all the GC candidates around the clump. 
It shows a clear radial central concentration, meaning that the GCs belong to this clump.
The central excess is seen out to $R_{\rm SW}\approx 9'$ ($=25$ kpc).
The blue GC clump is located at $R=19'$ so the background level of the clump is set to $N_{GC}(R=19')=0.6$ arcmin$^{-2}$. 
By integrating the background-subtracted radial profile, we obtain the total number of GCs, $N_{GC,SW}=33\pm5$. 
Then we can estimate the total mass of this substructure. 
Using the correlation for the dwarf ellipticals between $N_{GC}$ vs. $M_{dyn}$ of \citet{har13}, 
log$(N_{GC}) = 0.365 [$log$(M_{dyn}/M_\odot)- 9.2] + 1.274$, 
we obtain $M_{\rm dyn,SW}\simeq 7.4 \times 10^{9} M_\odot$. 
Because the 
SW BGC clump is the largest among the blue GC substructures, 
this value can be 
an upper limit of the mass of the progenitor which contributed to forming a stellar stream around M104.  
From the S\'ersic fitting result we find $R_{\rm e,GCS,SW}=4\farcm2\pm1\farcm2 \approx 8-15$ kpc.
This size is quite large compared to its dynamical mass. 
According to the relation of \citet{for18}, $M_{dyn}\approx 10^{12-13}M_\odot$ corresponds to $R_{e,GCS}\approx 10$ kpc. 
This indicates that it probably has been extended by tidal interactions with M104. 
The 
NE BGC clump is located too close to the galaxy center, so it is hard to estimate its size or mass. 

Additionally, we estimate the color of the south-western stellar stream. 
\citet{mar21} only had one broadband image so they could not measure the color of the stream. 
This stream is also seen in our MegaCam $g$ and $i$-band images so we can measure the $(g-i)$ color of the stream. 
The integrated magnitude of the stream is hard to estimate due to its faint and irregular morphology, so we estimate the surface brightness of the stream from the background level of the point sources on the stream. 
We estimate the surface brightness outside the stream to subtract the background effect. 
As a result, we obtain $(g-i)_0=0.75\pm0.05$. 

To verify the color estimation method, we additionally estimate the color of the two dwarf galaxies near the loop 
using the same method. 
The locations of these two dwarfs, NGC 4594-DGSAT1 and NGC 4594 DW1, are marked in Figure \ref{fig:spatial}. 
NGC 4594-DGSAT1 is one of the nearest dwarf galaxy from M104 and NGC 4594 DW1 is the second brightest dwarf galaxy near M104. 
\citet{car20} estimated the magnitude and 
color of the dwarf galaxies around M104 including these two dwarfs. 
We estimate the color of NGC 4594-DGSAT1 and NGC 4594 DW1 to be $(g-i)_0=0.65\pm0.05$ and $0.80\pm0.05$, respectively.
This is consistent with the result in \citet{car20}, $(g-i)=0.7\pm0.14$ and $0.8\pm0.01$, respectively. 
Therefore, our color estimation method is reliable. 
Thus we find that the stream color is very similar to the color of the two dwarf galaxies. 
These colors are also very similar to the color of the blue GCs. 
These dwarf galaxies are classified as dE type \citep{car20}, 
meaning that they are mainly composed of old stellar populations. 
From these results, we infer that the origin of the stellar stream and the SW BGC clump 
are dwarf galaxies. 

In conclusion, the progenitor of the SW stellar stream around M104 is probably a dwarf galaxy, which is consistent with the modeling result of \citet{mar21}. 
Still, there is a possibility that the blue GC clump is the result of projection effect \citep[see][]{hug22}. 
Spectroscopic confirmation for the GCs is needed to verify the clump and the progenitor of the stellar stream. 




\subsection{A Formation Scenario of M104} 
\subsubsection{GCs in M104}
Several results in this study support that the origin of the giant metal-poor outer halo of M104 is numerous minor mergers. 
First, the dominant population in the outer halo is blue (metal-poor) GCs. The progenitors of the blue GCs are mostly metal-poor dwarf galaxies. 
Second, the mass of the stream progenitor derived from the total number of GCs in the SW BGC clump 
is similar to the mass of dwarf galaxies.
Third, the color of the stream is similar to that of the early-type dwarf galaxies. 
Fourth, the red GC fraction is very low ($f_{RGC}=0.34\pm0.01$) which is similar to the fractions for the Virgo giant ETGs, and 
it is much lower than the fractions for the massive compact elliptical galaxies in clusters \citep[see Figure 12 of][]{kl21}.

\subsubsection{
No Metal-poor 
Stars at $R=33$ kpc in M104?} 
Another interesting but unresolved point in regard to the halo structure of M104, as suggested by \citet{coh20}, is that there are almost no metal-poor stars even in the outer region at $R=33$ kpc whereas metal-poor GCs dominantly exist. 
Our study shows that the metal-poor GCs are widely distributed from the center to the outer region at $R=100$ kpc, and the metal-poor GCs are dominant 
compared with the metal-rich GCs at $R>10$ kpc. 
This discrepancy between the distributions of metal-poor stars and metal-poor GCs is still hard to explain.



\subsection{GCLFs of M104 and GCLF calibration}


Previous estimates of the distance to M104 are well summarized in \citet{mcq16}. 
The most reliable distance to date is \citet{mcq16}'s distance, 
$(m-M)_0=29.90\pm0.03\pm0.07$ (9.55 Mpc), in the sense that they only use the halo RGB stars to derive the TRGB distance.
They measured the TRGB magnitude as F814W$_0=25.84\pm0.02$, 
and used a calibration of $M_{F814W}=-4.06\pm0.02\pm0.07$ mag.
The Extragalactic Distance Database \citep[EDD,][]{tul09} also estimates TRGB distance to M104, $(m-M)_0=29.85\pm0.03\pm0.07$ (9.33 Mpc). 
They measured the TRGB magnitude as F814W$_0=25.84\pm0.03$, $I_0=25.85\pm0.03$, 
and used a calibration of $M_I=-4.00\pm0.02$ mag.
The two distances are consistent within errors, and the main difference between the two is due to the difference in the calibration. 
We adopt \citet{mcq16}'s distance in this study. 


Using the TRGB distance, we can update the GCLF calibration. 
If we use the results of all GCs, we obtain GCLF TOM of 
$g_0^{\rm TOM} = 22.39\pm0.01$ and  
$i_0^{\rm TOM} = 21.64\pm0.01$.
Therefore, $M_g^{\rm TOM}=-7.51$ mag and $M_i^{\rm TOM}=-8.26$ mag.
If we use only the results from blue GCs, we obtain
$g_0^{\rm TOM}(BGC) = 22.28\pm0.02$ and  
$i_0^{\rm TOM}(BGC) = 21.59\pm0.02$.
Therefore, $M_g^{\rm TOM}(BGC)=-7.62$ mag and $M_i^{\rm TOM}(BGC)=-8.31$ mag.
Since we already checked that the V-band GCLF calibration of this study is consistent with the calibration based on the Milky Way GCs, we can tell that our new $g$ and $i$-band calibrations are reliable as well. 
Additional GCLF studies of other nearby galaxies with sufficient number of GCs will be helpful to calibrate GCLFs more reliably. 

\section{Summary and Conclusion} \label{sec:summary}

In this study, we obtain wide and deep images of M104 with CFHT/MegaCam observations and 
detect a large number of GCs to its outer region.
The color distribution of these M104 GCs shows two subpopulations: metal-poor GCs and metal-rich GCs.
From the analysis of their spatial distributions and their radial density profiles, we can conclude that M104 has a dual halo, 
which is similar to other massive ETGs. 

Primary results (and their implications) in this study are summarized as follows. 

\begin{enumerate}
    \item The radial extent of the GCs in M104 is found out to $R\approx 35'$ ($\sim100$ kpc), which is much farther than the previously known limit of $R\approx 20'$ \citep{har14}.
    The new boundary is also much larger than the distribution of diffuse galaxy light in \citet{bec84} and \citet{mal97a}.
    
    \item The GCs in M104 are composed of two subpopulations: blue GCs ($0.5<(g-i)_0<0.86$) with a peak metallicity [Fe/H]$=-1.4$ and red GCs ($0.86<(g-i)_0<1.2$) with a peak metallicity [Fe/H]$=-0.6$. 
    
    \item The radial number density profile and the color gradient profile of the GCs show a break at $R\approx 20'$ ($\sim 60$ kpc). 
    From this result, the M104 region is divided into three: the central region mainly composed of a bulge and a disk ($R\lesssim3'$), the inner region mainly composed of a metal-rich halo ($3' \lesssim R \lesssim 20'$), and the outer region mainly composed of a giant metal-poor halo ($20'\lesssim R \lesssim 35'$).
    
    \item The 
    low red GC fraction means that the origin of the giant metal-poor halo is dwarf satellites. 
    
    \item According to the analysis of 
    the SW BGC clump near the faint stellar light stream around M104, the progenitor of the stream is a dwarf galaxy with an upper mass of $M_{dyn,sub}\lesssim 10^{10}M_\odot$.

    \item M104 was formed as a classical massive ETG and then thought to have acquired its disk later as implied by \citet{dia18}'s simulation. 
    
    

    
    
\end{enumerate}

In the second paper in this series, we prepare the list of the GCs in M104 which were confirmed from MMT/Hectospec spectra, and we present a kinematic study of the confirmed GCs.

\acknowledgments
The authors are grateful to the anonymous referee for useful comments.
J.K. was supported by the Global Ph.D. Fellowship Program (NRF-2016H1A2A1907015) of the National Research Foundation.
This work was supported by the National Research Foundation grant funded by the Korean Government (NRF-2019R1A2C2084019). 
This work was supported by K-GMT Science Program (PID: 15AK06) funded through Korean GMT Project operated by Korea Astronomy and Space Science Institute (KASI).
We thank Brian S. Cho for his help in improving the English in this manuscript.

\facilities{CFHT(MegaCam)}

\end{document}